\newcommand{\diracslash}[1]{#1\llap{/\kern2pt}}

\newcommand{\be}{\begin{equation}}
\newcommand{\ee}{\end{equation}}
\newcommand{\bea}{\begin{eqnarray}}
\newcommand{\eea}{\end{eqnarray}}
\newcommand{\ba}[1]{\begin{array}{#1}}
\newcommand{\ea}{\end{array}}

\documentclass[12pt]{iopart}
\usepackage{graphicx}
\begin{document}

\title{Light scattering in Cooper-paired Fermi atoms}

\author{Bimalendu Deb}

\address{ Physical Research Laboratory, Navrangpura, Ahmedabad 380 009,
India}
\begin{abstract}
We present a detailed theoretical study of light scattering off
superfluid trapped Fermi gas of atoms at zero temperature. We
apply Nambu-Gorkov formalism of superconductivity to calculate the
response function of superfluid gas due to stimulated light
scattering taking into account the final state interactions. The
polarization of light has been shown to play a significant role in
response of Cooper-pairs in the presence of a magnetic field.
Particularly important is a scheme of polarization-selective light
scattering by either spin-component of the Cooper-pairs leading to
the single-particle excitations of one spin-component only. These
excitations have a threshold of $2\Delta$ where $\Delta$ is the
superfluid gap energy.  Furthermore, polarization-selective light
scattering allows for unequal energy and momentum transfer to the
two partner atoms of a Cooper-pair. In the regime of low energy
($<\!< 2\Delta$) and low momentum ($<2\Delta/(\hbar v_F)$, $v_F$
being the Fermi velocity) transfer, a small difference in momentum
transfers to the two spin-components may be useful in exciting
Bogoliubov-Anderson phonon mode. We present detailed results on
the dynamic structure factor (DSF) deduced from the response
function making use of generalized fluctuation-dissipation
theorem. Model calculations using local density approximation for
trapped superfluid Fermi gas shows that when the energy transfer
is less than $2\Delta_0$, where $\Delta_0$ refers to the gap at
the trap center, DSF as a function of energy transfer has reduced
gradient compared to that of normal Fermi gas.

\end{abstract}


\def\be{\begin{equation}}
\def\ee{\end{equation}}
\def\bearr{\begin{eqnarray}}
\def\eearr{\end{eqnarray}}
\def\zbf#1{{\bf {#1}}}
\def\bfm#1{\mbox{\boldmath $#1$}}
\def\hf{\frac{1}{2}}

\pacs{03.75.Ss,74.20.-z,32.80.Lg}

\maketitle

\section{Introduction}

Cold atoms are of enormous research interest in current physics.
The tremendous advancement in technology of cooling, trapping and
manipulation  \cite{cooling}  of atomic gases during 80's and 90's
has enabled researchers to achieve a low temperature down to a few
hundredth of a microKelvin. This led to the first realizations of
Bose-Einstein condensation (BEC) \cite{bec} in dilute gases of
ultracold bosonic atoms about a decade ago. Predicted in 1924 by
Einstein \cite{einstein} based quantum statistics of
indistinguishable particles discovered by Bose \cite{bose}, BEC in
gaseous systems had long been thought a subject of mere academic
pursuit beyond experimental reach because of the requirement of
ultralow temperature which was unimaginable even two decades ago.
The success in BEC is a breakthrough prompting researchers to look
for experimental realizations of many other theoretical
predictions of quantum physics using cold atoms. The most
remarkable property of such atoms is the tunability of the
atom-atom interaction over a wide range by an externally applied
magnetic field or other means. This provides an unique opportunity
to explore physics of interacting many-particle systems in a new
parameter regime. In this context, the focus of attention has been
now shifted to cold atoms obeying Fermi-Dirac statistics. Since
fermions are the basic constituents of matter,  research with
Fermi atoms under controlled physical conditions has important
implications in the entire spectrum of physical and chemical
sciences. In particular, it has significant relevance in the field
of superconductivity \cite{pred1,pred2}.

The quantum degeneracy in an atomic Fermi gas was first realized
by Jin's group \cite{jin} in 1999. Since then, cold Fermi atoms
have been in focus of research interest in physics today.  In a
series of  experiments, several groups
\cite{hulet,solomon,thomas,mit,italy,grimm} have demonstrated many
new aspects of degenerate atomic Fermi gases. In a remarkable
recent experiment, Ketterle's group \cite{ketterle} has realized
quantized vortices as a signature of Fermi superfluidity in a
trapped atomic gas.  Two groups-Innsbruck \cite{gap1} and JILA
\cite{gap2} have independently reported the measurement of pairing
gap in Fermi atoms.  Furthermore, Duke and Innsbruck groups
\cite{duke,innsbruck} have measured collective oscillations which
indicate  the occurrence of superfluidity \cite{stringari}. One of
the key issues in this field is the crossover
\cite{nozrink,randeria,crossover} between BCS state of atoms and
BEC  of molecules formed from Fermi atoms. Several groups have
demonstrated BEC \cite{molecules} of molecules formed from
degenerate Fermi gas. There have been many other experimental
\cite{expt} and theoretical investigations \cite{theory} revealing
many intriguing aspects of interacting Fermi atoms.

The analysis of response of Cooper-paired Fermi atoms due to
external perturbation (such as photon or rf field) is important
for understanding the nature of atomic Fermi superfluid.
 A method has been suggested to use
resonant light \cite{zoller} to excite one of the spin components
into an excited electronic state and thereby  making an interface
between normal and superfluid atoms. This is analogous to
superconductive tunnelling which has a threshold equal to the gap
energy $\Delta$. This has been recently implemented (albeit using
rf field) \cite{gap1,theogap1} to estimate gap energy. There have
been several other proposals \cite{zoller,huletp} for probing
pairing gap.

Our purpose here is to calculate response function of superfluid
Fermi gas due to stimulated light scattering that does not cause
any electronic excitation in the atoms. We particularly emphasize
the role of light polarization in single-particle excitations
which have a threshold $2\Delta$. We present a scheme by which it
is possible to have single-particle excitation in only one partner
atom (of a particular hyperfine spin state) of a Cooper-pair using
proper light polarizations in the presence of a magnetic field.
This may lead to better precision in spin-selective time-of-flight
detection of scattered atoms. Furthermore, spin-selective light
scattering allows for unequal energy and momentum transfer into
the two partner atoms of a Cooper-pair. This may be useful in
exciting Bogoliubov-Anderson (BA) phonon mode of symmetry breaking
by making small difference in momentum transfers received by the
two partner atoms from the photon fields. A number of authors
\cite{mottelson,griffin,minguzzi} have theoretically investigated
Bogoliubov-Anderson (BA)  mode \cite{bamode,anderson,martin} in
fermionic atoms as a signature of superfluidity.  BA mode
constitutes a distinctive feature of superfluidity in neutral
Fermi systems since it is associated with long wave  Cooper-pair
density fluctuations. However, experimental detection of this mode
is a challenging problem.

 We present a detailed theoretical
analysis of the response function of Cooper-paired atoms at zero
temperature due to light scattering. The stimulated light
scattering we discuss here is similar to Bragg spectroscopy used
by Ketterle's group for measuring structure factor of an atomic
BEC \cite{bragg}.  The response function we derive is applicable
for most general case of polarization-selective single-particle
excitations for unequal (or equal)  momentum as well as energy
transfers to the two partner atoms of a Cooper-pair. We develop
the theoretical framework for stimulated light scattering off
Cooper-paired Fermi atoms following the method used for describing
Raman scattering in superconductors \cite{abrikosov,klein}. We use
standard Nambu-Gorkov formalism of superconductivity
\cite{gorkov,nambu} to calculate the response function taking into
account the vertex correction due to final state interactions. We
deduce dynamic structure factor (DSF) from the response function
applying generalized fluctuation-dissipation theorem.  We present
detailed analytical and numerical results of our calculation of
DSF of trapped superfluid Fermi gas of atoms using local density
approximation. The inhomogeneity of trapped gas has a role in
distinguishing the DSF of superfluid gas from that of normal gas.
When the energy transfer is smaller than $2\Delta_0$ where
$\Delta_0$ is the gap at the trap center, the DSF of superfluid
gas as a function of energy transfer shows much reduced gradient
in comparison to that of normal gas. This is because of the fact
that the gap $\Delta$ has an inhomogeneous distribution gradually
vanishing at the edge of the trap.

The paper is organized in the following way. In the following two
sections, we define bare vertex in light scattering and response
function, respectively. In the fourth section, we discuss
stimulated light scattering in two-component $^6$Li Fermi atoms in
the presence of a magnetic field. We next describe in detail the
method of vertex correction in light scattering off Cooper-paired
Fermi atoms. In the sixth section, we discuss our analytical
results followed by description on numerical results in the
seventh section, and then we conclude.

\section{Bare vertex in light scattering}
To begin with, let us consider an elementary process of photon
scattering by a neutral atom. Let the atom's  initial and
scattered electronic state be denoted by $A$ and $B$,
respectively. The frequencies of the incident and scattered photon
are represented by $\omega_{1}$ and $\omega_{2}$, respectively.
According to second order perturbation theory,  the strength of
scattering is given by Kramers-Heisenberg formula \cite{sakurai}
\bearr \gamma_{BA} = \delta_{AB}\hat{\epsilon}_1\cdot
\hat{\epsilon}_2  -  \frac{1}{m_e\hbar}\sum_{I}   \left [
\frac{(\mathbf{p}.\hat{\epsilon}_2)_{BI}(\mathbf{p}.\hat{\epsilon}_1)_{IA}}{\omega_{IA}
- \omega_1} +
\frac{(\mathbf{p}.\hat{\epsilon}_1)_{BI}(\mathbf{p}.\hat{\epsilon}_2)_{IA}}{\omega_{IA}
+ \omega_2} \right ], \label{vert1} \eearr where $I$ denotes all
the intermediate atomic states  that can be coupled to the initial
and final atomic states $A$ and $B$ by the incident and scattered
photon fields. Here $\mathbf{p}$ and $m_e$ are the momentum and
mass of the valence electron of atom, $\hat{\epsilon}_{1(2)}$
denotes the polarization state of the incident (scattered) photon,
$\omega_{IA}$ is the atomic frequency between the states $I$ and
$A$. The atomic transition ($A\rightarrow B$) probability and the
differential scattering cross section of photons is proportional
to $|\gamma|^2$ \cite{sakurai}. It should be mentioned that
$\gamma$ does not depend on the momentum transfer $\mathbf{q}$
associated with the scattering, but it is sensitive to light
polarization directions. Let us now consider the particular case:
$A=B$ that is, before and after the scattering, the atom remains
in the same electronic state. Then, making use of the completeness
of the intermediate states $I$, one can rewrite the term
$\hat{\epsilon}_1.\hat{\epsilon}_2$ as \cite{sakurai} \bearr
\hat{\epsilon}_1.\hat{\epsilon}_2 = \frac{1}{m_e\hbar} \sum_{I}
\frac{1}{\omega_{IA}} \left [
(\mathbf{p}.\hat{\epsilon}_2)_{AI}(\mathbf{p}.\hat{\epsilon}_1)_{IA}
+
(\mathbf{p}.\hat{\epsilon}_1)_{AI}(\mathbf{p}.\hat{\epsilon}_2)_{IA}
\right ], \label{ee} \eearr Further, let us assume  $\omega_1
\simeq \omega_2 \simeq \omega_{IA}$, that is, the incident as well
as scattered light fields are in near resonance with the atomic
frequency. In such a case, the second term within the third
bracket on the right hand side (RHS) of Eq. (\ref{vert1}) is much
smaller than the first term,  because energy denominator of the
second term is of the order of optical frequency while that of the
first term can be chosen to be smaller by several orders of
magnitude. Thus, neglecting the second term, the bare vertex can
be written as \cite{sakurai} \bearr \gamma_{AA} =
-\frac{1}{m_e\hbar} \sum_{I} \frac{\omega_1
(\mathbf{p}.\hat{\epsilon}_2)_{AI}(\mathbf{p}.\hat{\epsilon}_1)_{IA}}{\omega_{IA}(\omega_{IA}
- \omega_1)}. \label{vert2} \eearr Next, using electric-dipole
approximation and the fact $\omega_1/\omega_{IA} \simeq 1$, one
can express \bearr \gamma_{AA} =  \Omega_0^{-1} \sum_{I}
\frac{(\mathbf{d}_{AI}.\hat{\mathcal{E}}_2)(\mathbf{d}_{AI}.\hat{\mathcal{E}}_1)}{\hbar^2(\omega_{AI}-
\omega_1)} \eearr where $\mathbf{d}_{AI}$ is the transition dipole
moment between the states $A$ and $I$, $\hat{\mathcal{E}}_i =
\mathcal{E}_i\hat{\epsilon}_i$ is the electric field and \bearr
\Omega_0 =
\frac{e^2\mathcal{E}_1\mathcal{E}_2}{m_e\sqrt{n_1}\hbar\omega_1\omega_2}
\eearr with $n_1$ being the number of incident photons.

\section{The response function}
To define response function of fermionic atoms due to an applied
laser field, we use the second-quantized operator
$a_{\sigma,\mathbf{k}}(a_{\sigma,\mathbf{k}}^{\dagger})$ which
describes the annihilation(creation) of an atom with hyperfine
spin $\sigma$ and center-of-mass momentum $\mathbf{k}$. These
operators satisfy fermionic algebra. The effective atom-field
hamiltonian is $H_{eff} = H_0 + H_I$, where \bearr H_0 =
\sum_{\sigma,\mathbf{k}} \hbar (\omega_k -
\delta)a_{\sigma,\mathbf{k}}^{\dagger}a_{\sigma,\mathbf{k}},
\hspace{0.4cm} \omega_k = \frac{\hbar k^2}{2m}, \eearr with
$\delta = \omega_1 - \omega_2$ being the frequency-difference
between incident and scattered photons. We assume that, except the
center-of-mass momentum, the spin or any other internal degrees of
atom does not change due to light scattering. By treating light
fields classically, the effective interaction hamiltonian can then
be written as
 \bearr H_{I} = \hbar\Omega_0 \sum_{\sigma,\mathbf{k}} \gamma_{\sigma\sigma}
a_{\sigma,\mathbf{k}+\mathbf{q}}^{\dagger}a_{\sigma,\mathbf{k}} +
{\mathrm H.c.} \eearr where $\mathbf{q}$ is the momentum
transferred to the atom due to photon scattering and
$\gamma_{\sigma\sigma}$ represents the bare vertex corresponding
to the ground hyperfine spin magnetic quantum number $\sigma$.

Now, one can define the density operators  by $\rho_q^{(0)} =
\sum_{\sigma,\mathbf{k}}
a_{\sigma,\mathbf{k}+\mathbf{q}}^{\dagger}a_{\sigma,\mathbf{k}}$
and \bearr
 \rho_q^{(\gamma)} = \sum_{k,\sigma}
 \gamma_{\sigma\sigma} a_{\sigma,\mathbf{k}+\mathbf{q}}^{\dagger}a_{\sigma,\mathbf{k}}
 \eearr
One can identify the operator $\rho_{q}^{(0)}$ as the Fourier
transform of the density operator in real space. The scattering
probability of incident particles (photons in the present context)
is related to the response or susceptibility \bearr
\chi(\mathbf{q},\tau-\tau') = -\langle
T_{\tau}[\rho_q^{(\gamma)}(\tau)\rho_{-q}^{(\gamma)}(\tau')]\rangle
\label{susctau} \eearr  of the target system by which the incident
particles are scattered. Here  $\langle \cdots \rangle$ means
thermal averaging and $T_{\tau}$ is the complex time $\tau$
ordering operator. The Fourier transform of this susceptibility is
\bearr \chi(\mathbf{q},\omega_n) = \frac{1}{2} \int_{-T}^T d\tau
e^{i\omega_n\tau} \chi(\mathbf{q},\tau) \label{susc} \eearr where
$T$ is the temperature and $\omega_n = 2\pi nT$ is the Matsubara
frequency with $n$ being an integer. The scattering cross section
is proportional to the generalized dynamic structure factor which
can be obtained by the generalized fluctuation-dissipation theorem
through the analytic continuation of $\chi(\mathbf{q},\omega_n)$
as \bearr S(\mathbf{q},\omega) = -
\frac{1}{\pi}[1+n_B(\omega)]\rm{Im}[ \chi(\mathbf{q},z=\omega +
i\delta)]. \label{flucdiss}\eearr We define the following
polarization matrix element: \bearr
\Pi_{ij}(\mathbf{q},\tau-\tau') = -\langle
T_{\tau}[\rho_q^{(i)}(\tau)\rho_{-q}^{(j)}(\tau')]\rangle \eearr
where $i,j \equiv \gamma,0$. The polarization bubble
$\Pi_{\gamma\gamma}$ is nothing but the susceptibility
$\chi(\mathbf{q},\tau-\tau')$ of Eq. (\ref{susctau}). The dynamic
structure factor is thus related to this polarization term by
fluctuation-dissipation relation as expressed in Eq.
(\ref{flucdiss}). The spectrum of density fluctuation is
proportional to the dynamic structure factor which can also be
defined as the Fourier transform of the two-time density-density
correlation function.

\section{stimulated light scattering in two-component Fermi atoms}

\begin{figure}
 \includegraphics{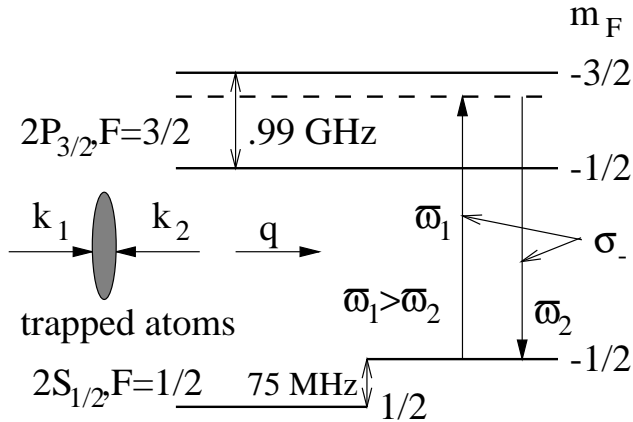}
 \caption{A schematic level diagram for polarization-selective light scattering
 in two-component Fermi gas of $^6$Li atoms}
 \label{diapol}
 \end{figure}
We would like to study stimulated light scattering in
two-component Fermi atoms. In particular, we consider trapped
$^6$Li Fermi atoms in their two lowest hyperfine spin states $\mid
g \rangle_1 = \mid 2{\rm S}_{1/2}, F = 1/2, m_{F} = 1/2 \rangle
\equiv \mid \uparrow \rangle $ and $\mid g \rangle_2 = \mid 2{\rm
S}_{1/2}, F=1/2, m_F = -1/2 \rangle  \equiv \mid \downarrow
\rangle$. For simplicity, the number of atoms in each spin
component is assumed to be the same. However, a mismatch in number
densities of the two spin components may lead to interior gap
superfluidity \cite{wilczek,deb1} in a Fermi gas of atoms. An
applied magnetic field tuned near the Feshbach resonance ($\sim
850$ Gauss) results in splitting between the two spin states by
$\sim 75$ MHz \cite{zwierlein}, while the corresponding splitting
between the excited states $ \mid e \rangle_1 = \mid 2{\rm
P}_{3/2}, F=3/2, m_{F} = -1/2 \rangle $ and $\mid e \rangle_2 =
\mid 2{\rm P}_{3/2}, F = 3/2, m_{F} = -3/2 \rangle $ is $\sim 994$
MHz \cite{thomas}.

Figure 1 shows the schematic level diagram for stimulated light
scattering by two-component $^6$Li atoms. Two off-resonant laser
beams with a small frequency difference are impinged on atoms, the
scattering of one laser photon is stimulated by the other photon.
In this process, one laser photon is annihilated and reappeared as
a scattered photon propagating along  the other laser beam. The
magnitude of momentum transfer is $q \simeq 2 k_L \sin(\theta/2)
$, where $\theta$ is the angle between the two beams and $k_L$ is
the momentum of a laser photon. Let both the laser beams be
$\sigma_{-}$ polarized and tuned near the transition $\mid
g\rangle_2 \rightarrow \mid e\rangle_2$. Then the transition
between the states $\mid g \rangle_1 $ and $\mid e \rangle_2 $
would be forbidden while the transition $\mid g \rangle_1
\rightarrow \mid e \rangle_1 $ will be suppressed due to the large
detuning $\sim 900$ MHz. This leads to a situation where the
scattered atoms remain in the same initial internal state $\mid g
\rangle_2$. Similarly, atoms in state $\mid g \rangle_1$ only
suffer scattering when two $ \sigma_{+}$ polarized lasers are
tuned near the transition $\mid g \rangle_1 \rightarrow \mid 2{\rm
P}_{3/2}, F = 3/2, m_F = 3/2\rangle$. Thus,  it is possible to
scatter atoms selectively of either spin components using
circularly polarized lasers in the presence of magnetic field.
Under such conditions, considering a uniform gas of atoms, the
effective laser-atom interaction Hamiltonian in electric-dipole
approximation can be written as
 \bearr
 H_I = \hbar \Omega_0 \sum_{{\mathbf k},\sigma=\uparrow,\downarrow}\gamma_{\sigma\sigma}
  a_{\sigma,{\mathbf k }+ {\mathbf q}}^{\dagger}
  a_{\sigma,\mathbf{
  k}} +
 {\mathrm H.c.} \label{eq2}
  \eearr
If $\sigma$ refers to $\mid \downarrow \rangle$ then  \bearr
\gamma_{\sigma\sigma} = \Omega_0^{-1} \sum_{i=1,2}
\frac{(\mathbf{d}_{22}.\hat{\mathcal{E}}_2)(\mathbf{d}_{22}.\hat{\mathcal{E}}_1)}{\hbar^2(\omega_{22}-
\omega_i)}. \label{gamma} \eearr where $d_{ii}$ is the transition
dipole matrix element between the ground $\mid g\rangle_i$ and the
excited $\mid e\rangle_i$ states. Similarly, if $\sigma$ is $\mid
\uparrow \rangle$ then the subscript ``22" should be replaced by
``11". In writing the above vertex term, we have also assumed that
both the laser beams are of almost equal intensity. For both the
laser beams having $\sigma_{-}$ polarization tuned near $\mid
g\rangle_2 \rightarrow \mid e \rangle_2 $ as in Fig. 1, one finds
$\gamma_{\downarrow\downarrow}>\!> \gamma_{\uparrow\uparrow}$.
 On the other hand, in the absence of magnetic field (or
in the presence of a weak magnetic field), the hyperfine magnetic
sub-levels of the ground and excited state would be degenerate (or
nearly degenerate). In such a case,  irrespective of whether both
the laser beams are unpolarized or equally polarized, we have
$\gamma_{\uparrow\uparrow} \simeq \gamma_{\downarrow\downarrow}$.

\section{light scattering in Cooper-paired
Fermi atoms: Vertex correction }

\begin{figure}
 \includegraphics[width=3.25in,height=2.15in]{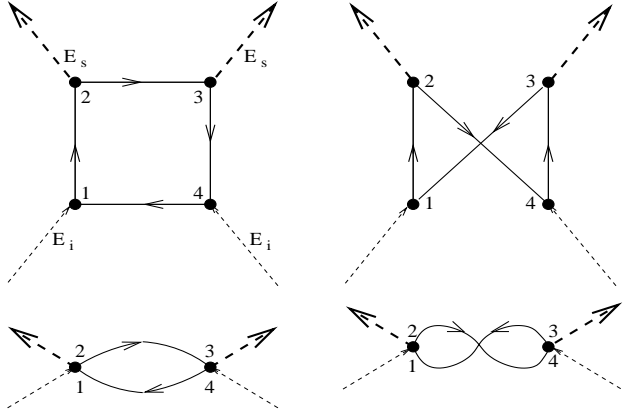}
 \caption{The upper part shows two irreducible four-vertex diagrams of stimulated light scattering off noninteracting normal Fermi gas of atoms.
 The thin dashed lines with arrows represent the incident laser field (with electric filed
 $\mathbf{E}_i$)
 and the thick dashed lines refer to the emitted photon stimulated by another laser field
 ($\mathbf{E}_s$). The
 operators $\mathbf{d}.\mathbf{E}_i$ and  $\mathbf{d}.\mathbf{E}_s$  act at
 the vertex pairs (1,4) and (2,3), respectively; where $\mathbf{d}$ is the transition dipole
 moment between the ground $\mid g_1 \rangle$ ($\mid g_2
 \rangle$) and the excited  $\mid e_1 \rangle$ ( $\mid e_2
 \rangle$ ) state. The  vertex pair (1,2) can be replaced
by an effective single vertex where the operator
$\gamma_{\sigma\sigma} \rho_q$ acts, where $\gamma_{\sigma\sigma}$
is given as in Eq. (\ref{gamma}). Similarly, the pair (3,4) can be
combined to form an effective vertex. Thus, the four vertex
diagrams effectively reduce to bubble diagrams as shown in the
lower part. Note that the role of incident and scattered fields
can be reversed, since an atom can absorb a photon from the laser
mode marked ``$\mathbf{E}_s$" and emit into the mode marked
``$\mathbf{E}_i$". By treating laser fields classically, the
effective vertex operators can be expressed only in terms of
atomic Fermi operators as in Eq. (\ref{eq2}). }
 \end{figure}

To study light scattering in Cooper-paired Fermi atoms, we apply
Nambu-Gorkov formalism that uses  the four Pauli matrices
\begin{eqnarray}\tau_0 = \left ( \begin{array}{cc}
 1 & 0 \\
0 & 1
\end{array} \right ), \hspace{0.25cm} \tau_1 = \left ( \begin{array}{cc}
 0 & 1 \\
1 & 0
\end{array} \right ) \hspace{0.25cm}
\tau_2 = \left ( \begin{array}{cc}
 0 & -i \\
i & 0
\end{array} \right ), \hspace{0.25cm} \tau_3 = \left ( \begin{array}{cc}
 1 & 0 \\
0 & -1
\end{array} \right )
\end{eqnarray}

The vertex equation is \cite{schrieffer} \bearr \Gamma(k_+,k_-) =
\tilde{\gamma} + i \int \frac{d^4k'}{(2\pi)^4} \tau_3
\mathbf{G}(k_+')
\Gamma(k_+',k_-')\mathbf{G}(k_-')\tau_3V(\mathbf{k},\mathbf{k}'),
\label{vertex} \eearr where $k_{\pm} = k \pm q/2$ and $k =
(\mathbf{k},k_0)$ is the
 energy-momentum  4-vector whose components are $k_3 = \xi_k$
 and $k_4=ik_0$.  In pairing
approximation, the Green function can be expressed in a matrix
form as \bearr
 G(k) = \frac{k_0 \tau_{0} + \xi_k\tau_3 +
 \Delta_k\tau_1}{k_0^2 - E_k^2 + i\delta}, \label{green}\eearr where $E_k =
 \sqrt{\xi_k^2 + \Delta_k^2}$ and $\xi_k = \epsilon_k-\mu$ with $\epsilon_k = \hbar^2 k^2/(2m)$. The bare vertex
 \bearr \tilde{\gamma} = \left( \begin{array}{cc}
 \gamma_{\uparrow\uparrow} & 0 \\
 0 & -\gamma_{\downarrow\downarrow}\end{array}  \right ). \eearr  Using Pauli matrices $\tau_0$ and
$\tau_3$, this can be rewritten as \bearr \tilde{\gamma} =
\gamma_0 \tau_0 + \gamma_3 \tau_3, \eearr where $\gamma_0 =
[\gamma_{\uparrow\uparrow} - \gamma_{\downarrow\downarrow}]/2$ and
$\gamma_3 = [\gamma_{\uparrow\uparrow} +
\gamma_{\downarrow\downarrow}]/2$. The susceptibility is given by

\bearr \chi(\mathbf{q},\omega) = \int
\frac{d^4k}{(2\pi)^4i}\rm{Tr}[\tilde{\gamma}_k\mathbf{G}(k_{+})\Gamma(k_+,k_-)\mathbf{G}(k_-)]
\eearr

\begin{figure}
 \includegraphics[width=3.25in]{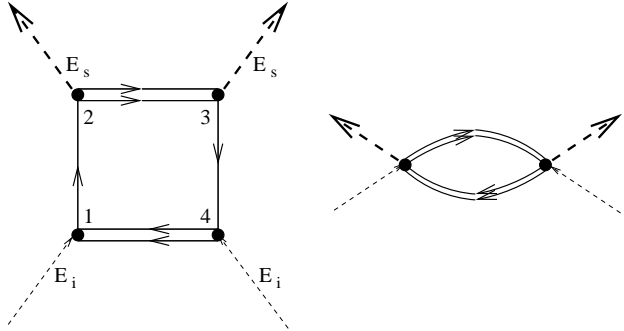}
 \caption{An irreducible four-vertex diagrams of stimulated light scattering in an atomic Fermi
 superfluid when the quasi-particles are assumed to be
 noninteracting. The double lines with arrows represent Nambu
 propagator for Cooper pairs. As in Fig 2, the four vertex diagram
 can be effectively represented by a two-vertex bubble diagram.}
 \end{figure}

\begin{figure}
 \includegraphics[width=3.25in]{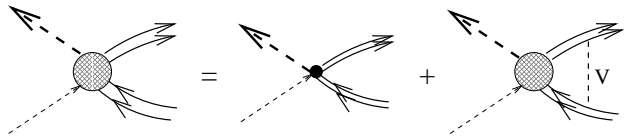}
 \caption{ Diagrammatic representation of vertex equation}
 \end{figure}

 \subsection{vertex equation and its solution}
 To solve the vertex equation, let us expand the vertex function
 in terms of Pauli matrices as
 \bearr
\Gamma(k_+,k_-) =
\sum_{i=0}^{3}\Gamma^{(i)}(\mathbf{k},\mathbf{q},\omega)\tau_i.
 \label{expansion}\eearr
Using Eqs. (\ref{green}) and (\ref{expansion}) in Eq.
(\ref{vertex}), we can write
 \bearr
\Gamma(\mathbf{k},\mathbf{q},\omega) = \tilde{\gamma}_k + i \int
\frac{d^4k'}{(2\pi)^4} V(\mathbf{k},\mathbf{k}') \frac{1}{(k_+'^2
+ \Delta_{k'}^2)(k_-'^2 + \Delta_{k'}^2)}\sum_{i=0}^{3}y_i\tau_i,
\label{vertexnew}\eearr where $k^2 = k_3^2 + k_4^2 = \xi_k^2 -
k_0^2$ and
 \begin{eqnarray} y_0 &=&
[k_{+0}^{\prime}\epsilon_{k_-'} +
k_{-0}^{\prime}\epsilon_{k_+'}]\Gamma^{(3)}
-i\Delta_{k'}[\epsilon_{k_+'} - \epsilon_{k_-'}]\Gamma^{(2)} +
\Delta_{k'}[k_{+0}'+k_{-0}']\Gamma^{(1)} \nonumber \\
&+& [\epsilon_{k_+'}\epsilon_{k_-'} + k_{+0}'k_{-0}' +
\Delta_{k'}^2]\Gamma^{(0)}, \nonumber
\end{eqnarray}
 \bearr y_1 &=&
-i[k_{+0}^{\prime}\epsilon_{k_-'} -
k_{-0}^{\prime}\epsilon_{k_+'}]\Gamma^{(2)}
-\Delta_{k'}[\epsilon_{k_+'} + \epsilon_{k_-'}]\Gamma^{(3)} -
\Delta_{k'}[k_{+0}'+ k_{-0}']\Gamma^{(0)} \nonumber \\ &+&
[\epsilon_{k_+'}\epsilon_{k_-'}-k_{+0}'k_{-0}'-\Delta^2]\Gamma^{(1)},
\nonumber  \eearr
 \bearr y_2 &=& - i\Delta_{k'}[k_{+0}^{\prime}
- k_{-0}^{\prime}]\Gamma^{(3)} -i\Delta_{k'}[\epsilon_{k_+'} -
\epsilon_{k_-'}]\Gamma^{(0)} -i [\epsilon_{k_+'}k_{-0}^{\prime} -
k_{+0}^{\prime}\epsilon_{k_-'} ]\Gamma^{(1)}  \nonumber \\ &+&
[\epsilon_{k_+'}\epsilon_{k_-'} - k_{+0}'k_{-0}' +
\Delta_{k'}^2]\Gamma^{(2)}, \nonumber  \eearr
 \bearr y_3 &=&  \Delta_{k'}[\epsilon_{k_+'}+ \epsilon_{k_-'}]\Gamma^{(1)}
 -i\Delta_{k'}[k_{+0}^{\prime} -
k_{-0}^{\prime}]\Gamma^{(2)}  +
 [\epsilon_{k_+'}\epsilon_{k_-'}
+ k_{+0}'k_{-0}' - \Delta_{k'}^2]\Gamma^{(3)} \nonumber \\ &+&
[\epsilon_{k_+'}k_{-0}^{\prime} +
k_{+0}^{\prime}\epsilon_{k_-'}]\Gamma^{(0)}.
 \nonumber
 \eearr
In writing the above equations, we have assumed $\Delta_{k_{\pm}}
\simeq \Delta_k$. Further, we can write $\epsilon_{k_{\pm}} =
\xi_k \pm \mathbf{v}_k.\mathbf{p}_q/2 + \epsilon_q$, where
$\mathbf{v}_k = \hbar\mathbf{k}/m$, $\mathbf{p}_q = \hbar q$ and
$\epsilon_q = p_q^2/(2m)$

Before performing the integration of Eq. (\ref{vertexnew}), we
note that the dominant contribution to the integral comes from
$k$-values near $\xi_k \simeq 0$, that is,
 $\hbar^2k^2/(2m) \simeq \mu$. Hence we can
approximate \bearr \int \frac{d^4\mathbf{k}}{(2\pi)^4}\simeq
\int\frac{d^3{\mathbf k}}{(2\pi)^3} \delta(\xi_k )\int \int
\frac{dk_3dk_0}{2\pi} \label{approx} \eearr where $k_3 \equiv
\xi_k $ denotes the third component of energy-momentum 4-vector $k
= (\mathbf{k},k_0)$. If the  potential $V(\mathbf{k},\mathbf{k}')$
is separable in two variables $\mathbf{k}$ and $\mathbf{k}'$, then
Eq. (\ref{vertex}) is analytically solvable. Let us, for
simplicity, replace $V(\mathbf{k},\mathbf{k}')$ by the well-known
mean field potential $V_{mf} = ga_s$ (where $g=4\pi\hbar^2/(2m)$)
which is expressed in terms of s-wave scattering length $a_s$. By
doing so, we are basically considering the weak-coupling case.
However, within mean-field approximation the strong-coupling limit
may be accessed by first  renormalizing the BCS mean-filed
interaction and then taking the limit $a_s \rightarrow \pm \infty$
as will be discussed later.

With the assumption of a $k$-independent gap $\Delta$, the double
integrations on $k_0$ and $k_3$ then resemble to those appearing
in relativistic equations in QED and so can be carried out
analytically by Feynman's method \cite{feynman}. The angular
integration is left to the last. There are basically two types of
integrals: \bearr I(q) = -i \int \frac{\Delta^2 dk_0dk_3}{(k_+^2 +
\Delta^2)(k_-^2 + \Delta^2)} \label{integral1} \eearr \bearr
I_{ij}(q) = -i \int \frac{(k_+)_i(k_-)_jdk_0dk_3}{(k_+^2 +
\Delta^2)(k_-^2 + \Delta^2)},  \hspace{0.2cm} i,j=3,4 \eearr These
integrals are explicitly calculated in Ref. \cite{vaks} using
Feynman's method of parametrization. For completeness, we here
reproduce the method of calculation. The  terms which are odd in
$k$ will not contribute to the integration and so those terms can
be omitted. Substituting  $k = \tilde{k} - (q/2 - qx)$ where $x$
is a parameter varying between 0 to 1, the integral of Eq.
(\ref{integral1}) can be reexpressed as \bearr I = -i \int_0^{1}dx
\int \frac{\Delta^2d\tilde{k}_0d\tilde{k}_3}{[\tilde{k}^2 +
\Delta^2 + q^2(x-x^2)]^2}. \label{integral2} \eearr The
$k_0$-integration can be carried out by residue method of complex
integration. The pole is $\tilde{k}_0 = \sqrt{\tilde{k}_3^2 + L}$,
where $L = \Delta^2 + q^2(x-x^2)$. Since $L$ has infinitesimally
negative imaginary part, the pole lies in the lower half of the
real axis. The residue is $-[4(\tilde{k}_3^2 + L)^{3/2}]^{-1}$.
After performing $\tilde{k}_3$- and $x$-integration , one obtains
the result $I(q)=f(q)/2$, where \bearr f(q) = \frac{\arcsin
\beta}{\beta\sqrt{1-\beta^2}}, \label{fbeta} \\
\nonumber \\
 \beta^2 =
\frac{\omega^2 - (\mathbf{v}_k.\mathbf{p}_q)^2}{4\Delta^2}
\label{beta}. \eearr The $k_3$-integration in Eq.
(\ref{integral2}) is divergent, therefore a cut-off frequency
$\omega_c$ is required as the upper limit of integration. After
having performed the integration, the vertex terms $\Gamma^{(i)}$
can be expressed as
 \bearr \Gamma^{(0)} &=& \gamma_0(k) -ga_s\int
 \frac{d^3\mathbf{k}}{(2\pi)^{3}}\delta(\xi_k) \nonumber \\ &\times& \left
 [\frac{\omega (\mathbf{v}_k.\mathbf{p}_q)(1-f)}{\omega^2-(\mathbf{v}_k.\mathbf{p}_q)^2}\Gamma^{(3)}
 -i\frac{\mathbf{v}_k.\mathbf{p}_q
 f}{2\Delta}\Gamma^{(2)} +
 \frac{(\mathbf{v}_k.\mathbf{p}_q)^2(1-f)}{\omega^2-(\mathbf{v}_k.\mathbf{p}_q)^2}\Gamma^{(0)}\right]
 \eearr
 \bearr \Gamma^{(1)} = -ga_s\int
 \frac{d^3\mathbf{k}}{(2\pi)^3}\delta(\xi_k)\left
 [\ln\frac{\omega_c}{|\Delta|}+(\beta^2-1)f\right]\Gamma^{(1)}
 \eearr
 \bearr \Gamma^{(2)} &=& -ga_s\int
 \frac{d^3\mathbf{k}}{(2\pi)^{3}}\delta(\xi_k) \nonumber \\ &\times& \left [-i\frac{\mathbf{v}_k.\mathbf{p}_q
 f}{2\Delta}\Gamma^{(0)}
+ \{\ln\frac{\omega_c}{|\Delta|}+\beta^2f\}\Gamma^{(2)}
-i\frac{\omega f}{2\Delta}\Gamma^{(3)}\right ] \label{gamma2}
\eearr
 \bearr
\Gamma^{(3)} &=&  \gamma_3(k) -ga_s\int
 \frac{d^3\mathbf{k}}{(2\pi)^{3}}\delta(\xi_k) \nonumber \\ &\times& \left [-i\frac{\omega f}{2\Delta}\Gamma^{(2)}
 +\frac{(\mathbf{v}_k.\mathbf{p}_q)^2-\omega^2 f}{\omega^2-(\mathbf{v}_k.\mathbf{p}_q)^2}\Gamma^{(3)}
 + \frac{\omega (\mathbf{v}_k.\mathbf{p}_q)(1-f)}{\omega^2-(\mathbf{v}_k.\mathbf{p}_q)^2}\Gamma^{(0)}\right ]\eearr

Since $\Gamma^{(1)}$ is decoupled from all other  vertex terms
including the bare ones ($\gamma_i$), we can set $\Gamma^{(1)} =
0$. Using the expansion of Eq. (\ref{expansion}), the
susceptibility can be written as
 \bearr \chi(\mathbf{q},\omega) &=& -
\frac{2(\Gamma^{(0)}-\gamma_0)}{ga_s}
 \gamma_0 + 2\int \frac{d^{3}{\mathbf
k}}{(2\pi)^{3}}\delta(\xi_k) \nonumber \\ &\times& \left [
\frac{(\mathbf{v}_k.\mathbf{p}_q)^2-\omega^2f
}{\omega^2-(\mathbf{v}_k.\mathbf{p}_q)^2}\Gamma^{(3)} -
\frac{i\omega f}{2\Delta}\Gamma^{(2)}+ \frac{\omega
(\mathbf{v}_k.\mathbf{p}_q)(1-f)}{\omega^2-(\mathbf{v}_k.\mathbf{p}_q)^2}\Gamma^{(0)}
\right]\gamma_3. \eearr We  note that the dressed part of
$\Gamma^{(0)}$ is proportional to the momentum transfer $q$,
therefore we have $\Gamma^{(0)} \simeq \gamma_0(k)$ in the low
momentum transfer regime, that is, for $q <\!< \xi^{-1}$, where
$\xi = \hbar v_F/(2\Delta)$ is the BCS coherence length.
Introducing the variable $ z = \cos\theta $, where $\theta$ is the
angle between $\mathbf{v}_k$ and $\mathbf{p}_q$, we can drop all
the terms odd in $z$ in the above equations, since upon
integration over $z$ those terms vanish. Thus  $\Gamma^{(0)}$ also
becomes decoupled while $\Gamma^{(2)}$ and $\Gamma^{(3)}$ form
only two coupled equations which can be analytically solved.

\subsection{gap equation}
The gap equation can be  obtained from Eq. (\ref{gamma2}) by
setting $\mathbf{q}$ and $\omega$ equal to zero and replacing
$\Gamma^{(2)}$ by the gap parameter $\Delta$. The resulting
equation reads \bearr \Delta  = -ga_s\int
 \frac{d^3\mathbf{k}}{(2\pi)^3}\delta(\xi_k)
\ln\frac{\omega_c}{|\Delta|}\Delta. \eearr The cut-off frequency
$\omega_c$  has been introduced ad-hoc to tackle the divergence
problem for the time being. This needs to be eliminated by the
method of regularization. To this end, we here recall that in
carrying out the various momentum integration, we made an
approximation: the integration was  restricted near the chemical
potential (which is nearly equal to Fermi energy in the weak
coupling regime). To restore the actual gap equation,  we here
remove this approximation and let $\omega_c \rightarrow \infty$
and thus obtain \bearr -\frac{1}{ga_s} = \frac{1}{2}\int
\frac{d^3\mathbf{k}}{(2\pi)^3} \frac{1}{\sqrt{\xi_k^2 +
\Delta^2}}. \label{unreggap} \eearr The gap defined by this
equation is however, divergent. To remove this divergence,  we
define regularized mean-field coupling
 by subtracting from the right hand side of Eq.
(\ref{unreggap}) the zero field contribution (i.e., $\Delta=0$ and
$\mu=0$). The resulting gap equation is \bearr -\frac{1}{ga_s}
=\frac{1}{2} \int \frac{d^3\mathbf{k}}{(2\pi)^3}
\left[\frac{1}{\sqrt{\xi_k^2 + \Delta^2}} -
\frac{1}{\epsilon_k}\right] \label{reggap} \eearr which yields
convergent results.  In the weak-coupling regime ($|a_s|k_F
<\!<1$), $\mu \simeq \epsilon_F \propto n^{2/3}$. The
strong-coupling regime ($|a_s|k_F
> 1$) may be accessed by simultaneously solving for the
interacting chemical potential $\mu$ from the single-spin  BCS
number-density equation \bearr n= \frac{1}{6\pi^2} k_F^3 =
\frac{1}{2} \int \frac{ d^3\mathbf{k}}{(2\pi)^3}\left (1-
\frac{\xi_k}{\sqrt{\xi_k^2 + \Delta^2}}\right ). \label{num}
\eearr This approach of solving the regularized gap plus the
number equation to access strong-coupling regime within the simple
mean-field framework fails to account for pairing fluctuation
effects which are particularly significant near $T_c$ in the
strong-coupling regime. However, far below $T_c$, the correction
due to the pairing fluctuation is very small as shown in Ref.
\cite{randeria}. The two coupled Eqs. (\ref{reggap}) and
(\ref{num}) admit analytical solutions  which are  obtained by
Marini {\it et al.} \cite{analyt} for the entire range of the
parameter $a_sk_F$ starting from weak interaction
($a_sk_F\rightarrow \pm 0$) to the unitarity limit
($a_sk_F\rightarrow \pm \infty$). In the unitarity limit, the
solutions  provide  $\mu = 0.59 \epsilon_F$ and $\Delta \simeq
1.16\mu$.  For convenience in solving the two coupled equations
numerically, we rewrite the equations  in terms of the two
dimensionless scaled variables $x=k/k_{\mu}$ and $y=\Delta/\mu$ as
\bearr \frac{2\pi}{k_{\mu}|a_s|}  = \int_{0}^{\infty} x^{2}\left [
\frac{1}{\sqrt{(x^2-1)^2 + y^2}} - \frac{1}{x^2} \right ] dx
\label{int1} \eearr

\bearr \left (\frac{k_F}{k_{\mu}} \right)^{3} = \frac{3}{2}
\int_0^{\infty} x^{2}\left [ 1 - \frac{x^2-1}{\sqrt{(x^2-1)^2 +
y^2}} \right ] dx \label{int2}  \eearr where $k_{\mu} =
\sqrt{2m\mu}/\hbar$. We have set $a_s = - |a_s|$. Calling the
right hand side of Eqs. (\ref{int1}) and (\ref{int2}) as $I_1$ and
$I_2$, respectively; eliminating $k_{\mu}$ from both the
equations, we obtain \bearr \frac{2\pi}{k_F|a_s|} =
\frac{I_1}{(I_2)^{1/3}}. \label{yeq} \eearr For given values of
the parameters $k_F$ and $|a_s|$, the  Eq. (\ref{yeq}) can be
solved for $y$. Then substituting this solution into Eq.
(\ref{int1}), one evaluates $\mu$ and so also the gap $\Delta =
\mu y$

\begin{figure}
 \includegraphics[width=3.25in]{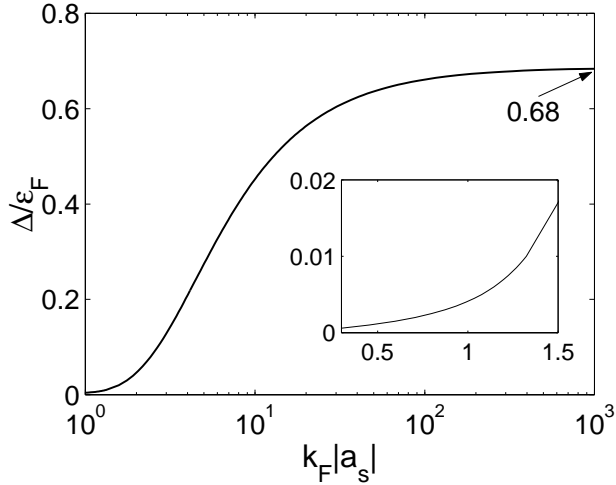}
 \caption{ Gap $\Delta$ (in unit of $\epsilon_F$) is plotted
 as a function of the dimensionless mean-field interaction parameter  $k_F|a_{s}|$ on semi-logarithmic scale.
 The inset shows
 the same plot for small interaction parameter on linear scale. In the limit $k_F|a_{s}|\rightarrow \infty$,
 the gap saturates at a value 0.68 $\epsilon_F$. }
 \label{figgap}
 \end{figure}

 \begin{figure}
 \includegraphics[width=3.25in]{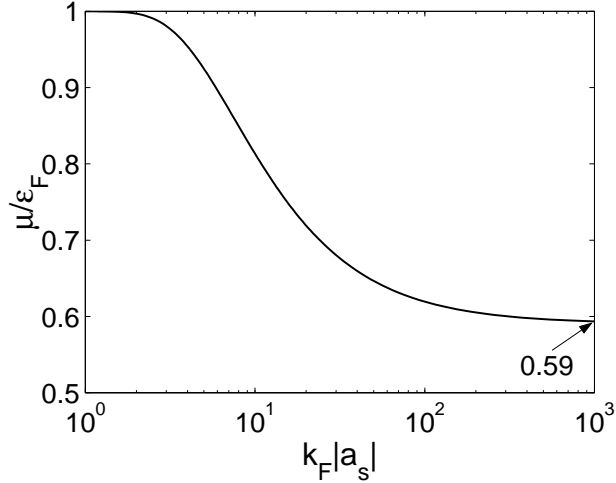}
 \caption{Chemical potential $\mu$ (in unit of $\epsilon_F$) is plotted
 against  parameter  $k_F|a_{s}|$. In the limit $k_F|a_{s}|\rightarrow \infty$,
 $\mu$  saturates at a value 0.59 $\epsilon_F$. In the limit $k_F|a_{s}|\rightarrow 0 $,
 $\mu$ goes to unity.}
 \label{figmu}
 \end{figure}

\subsection{solutions}

Now, to write down the  solutions of the various vertex terms
$\Gamma^{(i)}$  and the susceptibility $\chi$
 is  straightforward. Let $\kappa_s= N(0)ga_s$, where  $N(0) =
(\pi^2\hbar^2)^{-1}mk_F$ represents the single particle density of
states near the chemical potential. The various vertex terms can
be expressed as
 \bearr \Gamma^{(3)} =
\frac{\gamma_3}{1+\kappa_s F}, \eearr  \bearr \Gamma^{(2)} =
\frac{i\omega \langle f \rangle }{2\Delta\langle \beta^2 f \rangle
}\Gamma^{(3)}, \eearr and \bearr \Gamma^{(0)} =
\frac{\gamma_0}{1+\kappa_s\langle B \rangle}. \eearr
 Here \bearr F = \langle A \rangle +
\frac{\omega^2 \langle f \rangle^2}{ 4\Delta^2\langle \beta^2
f\rangle}, \label{F} \eearr  \bearr A =
\frac{(\mathbf{v}_k.\mathbf{p}_q)^2-\omega^2
f}{\omega^2-(\mathbf{v}_k.\mathbf{p}_q)^2}, \label{A} \eearr and
\bearr B =
\frac{(\mathbf{v}_k.\mathbf{p}_q)^2(1-f)}{\omega^2-(\mathbf{v}_k.\mathbf{p}_q)^2}.
\label{B} \eearr
 The symbol
$\langle X \rangle$ implies average of a function  $X$ over the
chemical potential surface:  $ \langle X \rangle = [N(0)]^{-1}\int
d^3\mathbf{k}\delta(\epsilon_k)  X$, since $X$ is an even function
of $z=\cos \theta$, we have $ \langle X \rangle = (1/2) \int_0^{1}
X(k_F,z) dz$. Making use of these vertex terms, the susceptibility
can be written as
 \bearr \chi(\mathbf{q},\omega) =
2N(0)\frac{\gamma_0^2\langle B \rangle}{1+\kappa_s\langle B
\rangle} + 2 N(0)  \left [\gamma_3^2F -
\frac{\kappa_s\gamma_3F^2}{1+\kappa_sF}\right ]
\label{chinew}\eearr
 We drop
the second term inside the third bracket which leads  to small
corrections due to Landau-liquid-like behavior without adding any
significant qualitative effect. Further, for $\kappa_s\langle
B\rangle <\!<1$, we have

  \bearr \chi(\mathbf{q},\omega) =
 2 N(0) \left [ \gamma_0^2 \langle B \rangle + \gamma_3^2 F
\right ] \label{chifinal}\eearr

\section{dynamic structure factor}
  The dynamic
structure factor is obtained from the response function $\chi$ via
analytic continuation of energy  $\omega \rightarrow \omega + i
0^+$. By means of generalized fluctuation-dissipation theorem as
embodied in  Eq. (\ref{flucdiss}), in the zero temperature limit
the dynamic structure factor is related to the imaginary part of
the density response function $\chi$  via  analytic continuation
of energy $\omega \rightarrow \omega + i 0^+$ as \bearr
S(\mathbf{q},\omega) = - \frac{1}{\pi}{\rm Im}[
\chi(\mathbf{q},\omega \rightarrow \omega + i 0^{+})].
\label{dsf}. \eearr The key function here is $f(\beta)$ of Eq.
(\ref{fbeta}), where $\beta$ is given in Eq. (\ref{beta}). As
$\omega \rightarrow \omega + i 0^+$, $\beta \rightarrow \beta + i
0^+$. We have the following analytic properties of $f(\beta)$:
 \bearr f(\beta) = h(\beta) +
\frac{i\pi/2}{\beta\sqrt{\beta^2-1}},
 \hspace{0.2cm} \beta > 1 \label{anal}\eearr
where \bearr h= -\frac{{\rm arcsinh} \sqrt{\beta^2-1}}{\beta
\sqrt{\beta^2-1}} \eearr

The use of Eqs. (\ref{F}), (\ref{A}) and (\ref{B}) in Eq.
(\ref{chifinal}) which, along with Eq. (\ref{anal}), on being
substituted in Eq. (\ref{dsf}) leads to the result \bearr
S(\mathbf{q},\omega) = \gamma_0^2 S_0(\mathbf{q},\omega) +
\gamma_3^2 S_3(\mathbf{q},\omega) \label{dsfnew} \eearr where
\bearr S_0(\mathbf{q},\omega) = N(0) \frac{\omega^2}{4\Delta^2}
\left \langle
\frac{(\mathbf{p}_q.\mathbf{v}_F)^2}{\beta^3\sqrt{\beta^2-1}}
\right \rangle \label{dsf1} \eearr \bearr S_3(\mathbf{q},\omega)
&=& N(0) \frac{\omega^2}{4\Delta^2}  \left \langle
\frac{1}{\beta^3\sqrt{\beta^2-1}} \right \rangle -N(0)
\frac{\omega^2}{4\Delta^2}\frac{1}{|\langle \beta^2
f \rangle|^2}\nonumber \\
&\times&  \left \{ \left \langle \frac{2\langle h \rangle \langle
\beta^2 h \rangle}{ \beta \sqrt{\beta^2-1} }\right \rangle -  {\rm
Re}[\langle f \rangle^2] \left \langle \frac{\beta }{
\sqrt{\beta^2-1}} \right \rangle \right \}  \label{dsf3} \eearr

\section{Analytical results and discussions}
Equation (\ref{dsfnew}) gives an expression for dynamic structure
factor of a homogeneous  Fermi superfluid when the excitations are
of single-particle type  for the parameters satisfying $\beta >
1$. Different amount of energy transfers (or excitations) to the
two constituent partners  of a broken Cooper-pair can be made  by
appropriately selecting the polarization states of the exciting
two laser beams and tuning their frequency from the excited atomic
state in the presence of a magnetic field. This fact is taken into
account in the expression of (\ref{dsfnew}), because any nonzero
value of
 the term $\gamma_0$  means unequal excitation of the two partners.
 For instance, two extreme cases
 can be mentioned: Case-I: For unpolarized light
 in the absence of magnetic field, equal amount of energy transfer
 occurs to the two partners resulting in $\gamma_0=0$; Case-II: On the
 other hand, for circularly polarized light in the presence of
 strong magnetic field,
we have $\gamma_0^2 \simeq \gamma_3^2$ meaning only either
 partner can be excited. We will present our numerical results for
 these two extreme cases. To compare our results with the known results
 for normal Fermi system in the limit $\Delta \rightarrow 0$, we
 will use in Case-I the limit $\gamma_{\uparrow\uparrow} \simeq \gamma_{\downarrow\downarrow} \rightarrow 1$
 meaning $\gamma_0\rightarrow 0$ and $\gamma_3  \rightarrow 1$. In
 Case-II, we will use the limit $\gamma_{\uparrow\uparrow} \simeq 0$ $\gamma_{\downarrow\downarrow} \rightarrow 1$
implying that $\gamma_3=-\gamma_0 \rightarrow 1/2$. Intuitively,
one may understand that the Case-II would be significantly
different from Case-I both qualitatively and quantitatively. In
the Case-II,  upon receiving an energy $\omega$ ($>2\Delta$)  from
an incident photon, one partner of a Cooper-pair moves out of  the
Fermi sphere, while the other partner remains within the Fermi
sphere. Let us consider an elementary process of single photon
scattering by a Cooper-pair. Suppose, the Cooper-pair consists of
an atom A having spin $\downarrow$ and momentum $\mathbf{k}$ and
another atom B with spin $\uparrow$ and momentum $-\mathbf{k}$.
When this Cooper-pair is broken due to stimulated scattering of
$\sigma_{-}$ polarized photon in a situation like Case-II, atom A
will move out of the Fermi surface as an excited quasi-particle
with momentum $\mathbf{k}+\mathbf{q}$ with certain probability
given by BCS correlation and atom B will have certain probability
of remaining within the Fermi sphere moving as a quasi-particle
with momentum $-\mathbf{k}$. Thus, only one partner of the
Cooper-pair will contribute to the intensity of scattered atoms
reducing the strength of the density fluctuation spectrum compared
to that of Case-I. However, there could be some advantage in
detecting the scattered atoms in Case-II by spin-selective
time-of-flight measurement technique as we will discuss later in
the concluding section.

\subsection{Case-I: Leading approximations} In this case, we have $\gamma_0 = 0$. In the limit
$\gamma_3 \rightarrow 1$, \bearr S_I(\mathbf{q},\omega) =
 S_3(\mathbf{q},\omega) \eearr which is given by Eq.
(\ref{dsf3}).  For $\beta
>1$, in the leading approximation in terms of $\beta^{-1}$, this
 reduces to the
form \bearr S_I^{\rm{lead}}(\mathbf{q},\omega) =N(0)
\frac{\omega^2}{4\Delta^2}\left \langle
\frac{1}{\beta^3\sqrt{\beta^2-1}} \right \rangle, \hspace{0.2cm}
\beta >1 \label{leadapp}\eearr which is devoid of any vertex
correction. The same expression can be derived by taking
$\Gamma^{(3)} \rightarrow \gamma_3$, $\Gamma^{(0)} \rightarrow
\gamma_0$ and $\Gamma^{(2)} \rightarrow 0$  meaning that we use
bare vertex only. This is also obtainable from  the static BCS-
Bogoliubov mean-filed treatment as shown in the appendix.  Because
of the absence of vertex correction,  it violates the Ward
identities \cite{ward} that guarantee the conservation of total
particle number and the obeyance of the continuity equation.

To perform the integration over $z$ in Eq. (\ref{leadapp}), it is
convenient to change the variable into \bearr x = \frac{p_q v_k
z}{\sqrt{\omega^2 - 4\Delta^2}} \label{x} \eearr The condition
$\beta
>1$ implies $x<1$. Then the Eq. (\ref{leadapp}) can be expressed
as \bearr S_I^{\rm{lead}}(\mathbf{q},\omega) =
\frac{2N(0)\Delta^2}{\omega p_q v_F} \int_0^{x_0}dx
\frac{1}{(1-jx^2)^{3/2}\sqrt{1-x^2}} \nonumber \\
\label{leadint}
\eearr where $j = 1 - 4\Delta^2/\omega^2$ and \bearr x_0 = {\rm
Min}\left [1,\frac{p_qv_F}{\sqrt{\omega^2 - 4\Delta^2}}\right ].
\eearr
 For $2\Delta <\omega < \sqrt{(p_qv_F)^2 + 4\Delta^2}$, we have
 $x_0 =1 $ and the result is
 \bearr S_I^{\rm{lead}}(\mathbf{q},\omega) =
 \frac{N(0)\omega}{2p_qv_F} E(j), \label{dsfnorm} \eearr where $E(j)$ is the complete
 elliptic integral. Note that in the limit $\Delta \rightarrow 0$
 $S_I^{\rm{lead}}(\mathbf{q},\omega)$ reduces to the form
$N(0)\omega/(2p_qv_F)$ which is same as that of a normal quantum
fluid of noninteracting quasi-particles within the energy range
$0< \omega < v_F p_q$ \cite{nozpines}. The dynamic structure
factor reaches a maximum at $\omega_0=\sqrt{(p_qv_F)^2 +
4\Delta^2}$. As $\omega$ increases above $\omega_0$, $x_0$
decreases below unity and hence the integral in Eq.
(\ref{leadint}) decreases.

In view of the forgoing analysis, we now verify how far  f-sum
rule is fulfilled by the dynamic structure factor as given by Eq.
(\ref{leadapp}). To this end, we separate the integral over energy
in the sum rule  \bearr \int \omega S(\mathbf{q},\omega)d\omega =
\int_0^{\omega_0}\cdots d\omega + \int_{\omega_0}^{\infty} \cdots
d\omega \label{sumrule}.  \eearr Since Eq. (\ref{leadapp}) holds
good for $\omega>2\Delta$, the first integral appearing on the
right hand side of  Eq. (\ref{sumrule}) results in \bearr I_{1} =
\int_{2\Delta}^{\omega_0}\omega S(\mathbf{q},\omega)d\omega =
\frac{N(0)}{2p_qv_F} \int_{2\Delta}^{\omega_0}\omega^2E(j)d\omega
\eearr where we have used the Eq. (\ref{dsfnorm}). In the limit
$\Delta\rightarrow 0$, $E(j)\rightarrow 1$ and so we obtain \bearr
I_1=\frac{N(0)(p_qv_F)^2}{6} =
\frac{k_F^3}{3\pi^2}\frac{p_q^2}{2m}=\frac{Np_q^2}{2m} \eearr
where we have used $N(0)=(\pi\hbar)^{-2}mk_F$. Here $N$ represents
the total number of particle per unit volume.  The second integral
on the RHS of Eq. (\ref{sumrule}) is much smaller than the first
one. Thus, we find that in the limit $\Delta\rightarrow 0$, or
alternatively, for $p_qv_F>\!>2\Delta$ and $\omega >\!> 2\Delta$,
that is, for large momentum and energy transfer, the dynamic
structure factor as given by Eq. (\ref{leadapp}) approximately
satisfies the f-sumrule.  In this context, it may be worthwhile to
mention here that for evaluating gap energy from the measurements
of the scattering cross section of the light-scattered atoms
released from a trap, large momentum transfer is indeed required
to distinguish the scattered atoms from the un-scattered ones
\cite{deb1,deb2}.  For single-particle excitation ($\beta>1$) with
small energy transfer, this leading approximation is not valid and
the second term on the RHS of Eq. (\ref{dsf3}) makes significant
contribution resulting from vertex correction. We will show in the
appendix that the DSF in leading order approximation is obtainable
from BCS-Bogoliubov mean-field treatment that does not take into
account final state (quasi-particles) interaction.

\subsection{Case-II: Leading approximations} In this case $\gamma_{\uparrow\uparrow}\rightarrow 0$
and  $\gamma_{\downarrow\downarrow}\rightarrow 1$ implying
$\gamma_0^2 \simeq \gamma_3^2 \ne 0$. Let us use $\gamma_0^2
\simeq \gamma_3^2 \rightarrow 1/4$. Then we have \bearr
S_{II}(\mathbf{q},\omega) = \frac{S_0(\mathbf{q},\omega) +
S_3(\mathbf{q},\omega)}{4}. \eearr  The angular integration in Eq.
(\ref{dsf1}) can be conveniently performed using the $x$-variable
as already introduced in Eq. (\ref{x}). Explicitly, this takes the
form \bearr S_0(\mathbf{q},\omega) = \frac{2N(0)\Delta^2}{\omega
p_q v_F} \int_0^{x_0}dx \frac{j x^2}{(1-jx^2)^{3/2}\sqrt{1-x^2}}
 \eearr where $j$ and $x_0$ are already defined
above. For $2\Delta <\omega < \sqrt{(p_qv_F)^2 + 4\Delta^2}$,
$x_0=1$ and the result is \bearr
 S_0(\mathbf{q},\omega)=\frac{2N(0)\Delta^2}{\omega p_q v_F}
\frac{\pi j}{4} \hspace{0.05cm}_2F_1(3/2,3/2;2,j) \label{z1}
\eearr $_2F_1(a,b;c,d)$ is the hypergeometric function. In the
limit $\Delta \rightarrow 0$, $\frac{\pi j(1-j)}{4}
\hspace{0.05cm}_2F_1 \rightarrow 1$, hence we obtain \bearr
S_0(\mathbf{q},\omega) = \frac{N(0)\omega}{2 p_q v_F} \eearr which
again coincides with the form of the DSF of normal fluid within
the specified parameter regime.

In passing, we reemphasize  that the leading order approximations
are valid for $\beta
>\!>1$, that is, for large energy transfer. In this limit,
$j\rightarrow 1$ and so  DSF's in both the cases  tend to become
equal. All he results followed from leading order approximations
can also be obtained with BCS-Bogoliubov mean-field approximation
without any vertex correction as illustrated in the appendix.

\subsection{Bogoliubov-Anderson mode}
Now let us consider the case $0 \le \beta <\!< 1$, that is
$\mathbf{v}_k.\mathbf{p}_q \le \omega <\!< 2\Delta$. In this case,
the second term in Eq. (\ref{F}) dominates over all other terms.
This term leads to Bogolibov-Anderson collective phonon mode
appearing as a pole in $\chi$. It is evident that the origin of
this pole lies in the vertex correction, since this is also the
pole of $\Gamma^{(2)}$. The pole is given by \bearr \langle
[\omega^2 - (\mathbf{v}_F.\mathbf{p}_q )^2]f \rangle = 0. \eearr
In the limit $q \rightarrow 0$ and $\omega \rightarrow 0$, $f
\simeq 1$ and
 hence the pole is \bearr \omega_{\rm{BA}} =
\frac{1}{\sqrt{3}}v_Fp_q \eearr The BA mode restores the
continuous symmetry which is broken by BCS ground state. It is
required to fulfill the Ward identities \cite{ward}. In the low
momentum and low energy limit ($0 \le \beta <\!< 1$)
 the dynamic structure factor can be obtained by
linearizing  the denominator of the second term in Eq.
(\ref{chifinal}) around the BA mode. By approximating $f \simeq
1$, we then obtain \bearr S(\mathbf{q},\omega) = N(0)\gamma_3^2
\frac{\omega^2}{2\omega_{BA}}\delta(\omega - \omega_{BA}).
\eearr With $\gamma_3\rightarrow1$,  this satisfies the $f-$sum
rule \bearr \int_0^{\infty} \omega S(\omega,\mathbf{q}) d\omega =
\frac{Nq^2}{2m} \eearr where $N$ is the total number of particles.

To have higher order (in terms of $\xi q$) corrections
\cite{mancini} to the BA mode, we expand the function $f(\beta)$
to the fourth order in $\beta$ and obtain the result \bearr
f(\beta) \simeq 1 + \frac{2\beta^2}{3} + \frac{\beta^4}{12} \eearr
Then the pole is then given by \bearr \langle \beta^2 f(\beta)
\rangle \simeq \langle \beta^2 + \frac{2\beta^4}{3} \rangle =0
\eearr resulting in  \bearr \omega_{BA}^2 =
\frac{(v_Fp_q)^2}{3}\left
[1-\frac{8}{45}\left(\frac{v_Fp_q}{2\Delta}\right)^2\right]
\label{bamode} \eearr BA mode is well defined in the low momentum
regime, i.e., for $\xi q=v_Fp_q/(2\Delta) <\!<1$. For large
momentum, it becomes ill defined due to Landau damping. To get the
dynamical correction to BA mode \cite{anderson,ambe}, the right
hand side of Eq. (\ref{bamode}) needs to be multiplied by a factor
$[1-g|a_s|N(0)]$.

Before closing this section, we would like to stress that the
polarization-selective small angle stimulated light scattering may
be useful in exciting BA mode. Because, unequal momentum and
energy transfer can be accomplished by making
$\gamma_{\uparrow\uparrow}\ne \gamma_{\downarrow\downarrow}$. This
will lead to unequal response of the two spin states. In the small
momentum and energy transfer regime, this will result in large
wave-length  center-of-mass motion of Cooper-pairs and hence
superfluid density fluctuation \cite{martin}. However, how to
detect this BA mode of superfluid trapped atoms is presently
unknown.

\section{Numerical results and discussion}

\begin{figure}
 \includegraphics[width=3.25in]{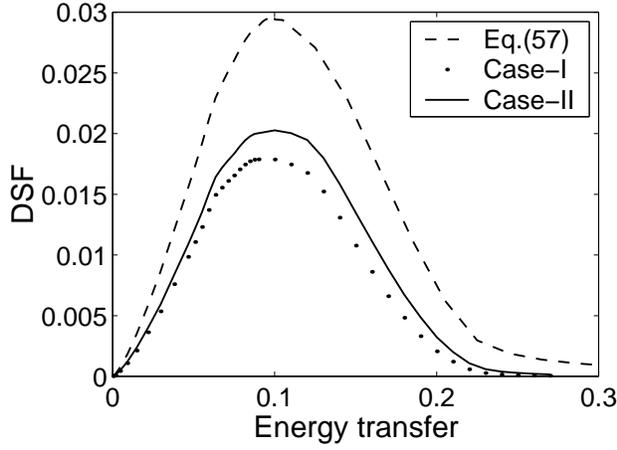}
 \caption{Scaled dimensionless dynamic structure factor (DSF) of
 superfluid trapped Fermi gas
 as a function of dimensionless energy transfer ( $\omega/\epsilon_F$)
  for different cases (see the text). For the sake of comparison, DSF for the two cases are
 scaled differently: $S_I(\omega,\mathbf{q})$ and
 $S_{II}(\omega,\mathbf{q})$ are scaled by the factors $1/[2N(0)]$
 and $1/N(0)$, respectively. The momentum transfer is kept fixed
 at $q/k_F = 0.2$.
 The scattering length is
 $a_s = 0.51 k_F^{-1}$ for which the BCS gap
 Eq. (\ref{bcsgap}) yields  the value of the gap at the trap center
 as $\Delta_0 = 0.05 \epsilon_F$. Case-I
 (dotted)
 refers to the unpolarized light in the absence of magnetic field,
case-II refers to the circularly polarized light in the presence
of magnetic field. Dashed curve (Eq. (57)) is for unpolarized
light without vertex correction when only BCS-type mean-field is
used.}
 \label{fign1}
 \end{figure}

 \begin{figure}
 \includegraphics[width=3.25in]{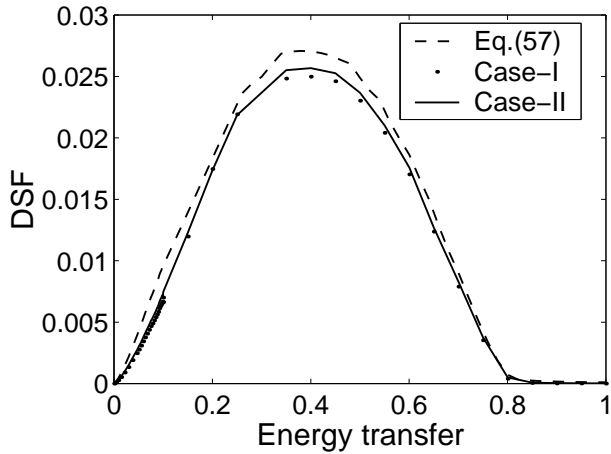}
 \caption{Same as in Fig. 1 but for $q = 0.8 k_F$ }
 \label{fign2}
 \end{figure}

We now apply the formalism discussed above to  harmonically
trapped superfluid Fermi atoms. For simplicity, we consider an
isotropic optical trap characterized by the length scale $a_{ho}=
\sqrt{\hbar/(m\omega_{ho})}$, where $\omega_{ho}$ is the trapping
frequency.   In Thomas-Fermi local density approximation (LDA)
\cite{houbiers},  the state of the system is governed by \bearr
\epsilon_F({\mathbf r})+ V_{ho}({\mathbf r}) + U({\mathbf r}) =
\mu, \label{eqstate} \eearr where $\epsilon_F ({\mathbf r}) =
\hbar^2 k_F({\mathbf r})^2 /(2m)$ is the local Fermi energy,
$k_F({\mathbf r})$ denotes the local Fermi momentum which is
related to the single-spin local number density by $n({\mathbf r})
= k_F({\mathbf r})^3/(6\pi^2)$. Here $U$ represents the mean-field
interaction energy and $\mu$ is the chemical potential. At low
energy, the mean-field interaction energy depends on the two-body
s-wave scattering amplitude $f_0(k)= -a_s/(1+ia_sk)$, where $a_s$
represents  s-wave scattering length and $k$ denotes the relative
wave number of two colliding particles. In the dilute gas limit
($|a_s|k <\!<1$), $U$ becomes proportional to $a_s$ in the form
$U({\mathbf r}) = \frac{4\pi\hbar^2 a_s}{2m} n({\mathbf r})$.
 In the unitarity limit $|a_s|k \rightarrow \infty$,
the scattering amplitude $f_0 \sim i/k$ and hence  $U$ becomes
independent of $a_s$. It then follows from a simple dimensional
analysis that in this limit, $U$ should be proportional to the
Fermi energy: $U({\mathbf r}) = \beta_{u} \epsilon_F({\mathbf r})$
where $\beta_{u}$ is the constant.  In this limit, the pairing gap
also becomes proportional to the Fermi energy. Based on the
regularized mean-field approach discussed earlier and LDA, the
zero-temperature density profiles \cite{strinati}, momentum
distribution \cite{pitaevskii} and  the finite temperature effects
\cite{perali} of superfluid trapped Fermi atoms have  been
recently studied. For dilute gas limit, the local density
distribution of trapped gas may be may be approximated by
neglecting the interaction term $U$ in Eq. (\ref{eqstate}). In the
BCS limit $(k_F a_s \rightarrow 0^{-}$), the gap is exponentially
small and can be expressed by the well known formula \bearr
\Delta_{\rm BCS} \simeq \frac{8 \epsilon_F}{e^2}\exp\left
[-\frac{\pi}{2k_F |a_s|}\right ], \label{bcsgap} \eearr where
$\epsilon_F$ is the Fermi energy.

\begin{figure}
 \includegraphics[width=3.25in]{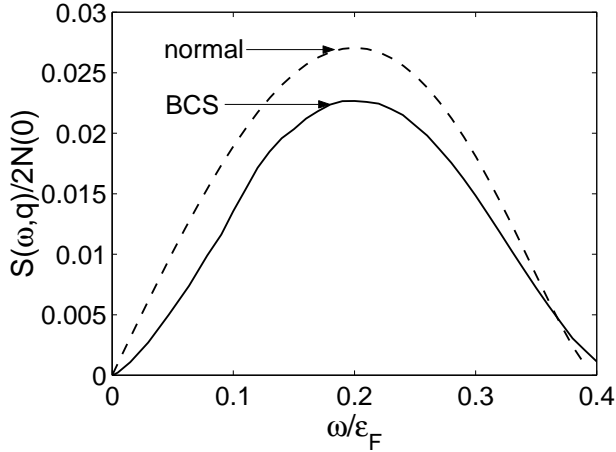}
 \caption{DSF of  BCS superfluid (solid) and
 normal fluid (dashed) are plotted as a function of energy
 transfer for $q = 0.4 k_F$ and $\Delta_0 = 0.05$ for  the case-I.}
 \label{fign3}
 \end{figure}

Under LDA,  the density profile of a trapped Fermi  gas is given
by \bearr n({\mathbf r}) = n({\mathbf 0}) (1 - r^2/R_{TF}^2
)^{3/2}, \label{nr} \eearr where $ n({\mathbf 0}) =
1/(6\pi^2\hbar^3)[2m\mu/(1+\beta_{u})]^{3/2}$ is the density of
the atoms at the trap center. Here $R_{TF}^2 =
2\mu/(m\omega_{ho}^2)$ is the Thomas-Fermi radius of the trapped
atomic gas. The normalization condition on Eq. (\ref{nr}) gives an
expression for $
 \mu = (1+\beta_u)^{1/2} (6N_{\sigma})^{1/3}\hbar\omega_{ho}
$ where $N_{\sigma}$  is the total number of atoms in the
hyperfine spin $\sigma$. The Fermi momentum at the trap center
$k_F^0 = [3\pi^2 n({\mathbf 0})]^{1/3} = (1+\beta_u)^{-1/4}k_F$
where $ k_F = (48N_{\sigma})^{1/6}/a_{ho} $ is the Fermi momentum
of the noninteracting trapped gas. Under LDA, the dynamic
structure factor is given by \bearr S(\mathbf{q},\omega) =
\frac{1}{V_{TF}} \int d^3{\mathbf r} S_r(\mathbf{q},\omega) \eearr
where $S_r(\mathbf{q},\omega)$ is the DSF for Fermi momentum
$k_F({\mathbf r})$ evaluated at a position $\mathbf{r}$ assuming
the system is locally uniform. Here $V_{TF} = (4/3) \pi R_{TF}^3$
is the Thomas-Fermi volume.

\begin{figure}
 \includegraphics[width=3.25in]{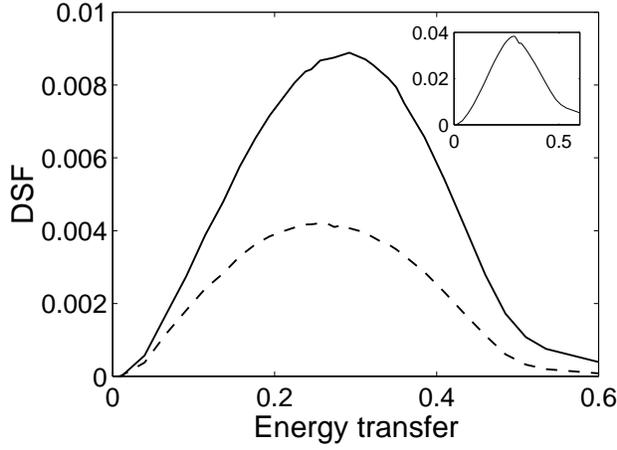}
 \caption{Dimensionless scaled DSF $S_I(\omega,\mathbf{q})/(2N(0))$ (dashed) and $S_{II}(\omega,\mathbf{q})
 /N(0)$ Vs. dimensionless energy
 transfer $\omega/\epsilon_F$ for $q = 0.4 k_F$ and
 $a_s=3.21k_F^{-1}$. For this larger scattering length,
 using the regularized gap Eq. (\ref{reggap}), the gap  is calculated to be  $\Delta_0 = 0.15 \epsilon_F$.
 The inset shows the corresponding plot in Case-I for leading order approximation as given  Eq. (57). }
 \label{fign4}
 \end{figure}

In Figs. \ref{fign1}-\ref{fign4}, we show DSF (calculated using
LDA) of superfluid trapped Fermi gas  due to single-particle
excitations only as a function of energy transfer under different
physical conditions. Figs. \ref{fign1}-\ref{fign3} are plotted for
$a_sk_F<1$ with gap given by BCS gap Eq. (\ref{bcsgap}) while Fig.
\ref{fign4} is for $a_sk_F>1$ with gap determined by the
regularized gap Eq. (\ref{reggap}) coupled with the superfluid
number  Eq. (\ref{num}). For the sake of better comparison, we
have scaled the DSF in Case-II (denoted by $S_{II}$ hereafter) by
a factor $[N(0)]^{-1}$ while DSF in Case-I (denoted by $S_I$
hereafter) is scaled by half of this factor, that is,  by
$[2N(0)]^{-1}$. Here $N(0)$ refers to the density of states at the
trap center. By comparing Fig. \ref{fign1} and Fig. \ref{fign2}
which are plotted for lower and higher momentum transfer,
respectively, we infer that the vertex correction is most
significant in low momentum and energy transfer regime. At high
momentum and energy transfer regime, mean-field approximation
seems to be reasonably good. Furthermore, at lower momentum
transfer,  $S_{II}$ shows larger deviation from $S_I/2$ with both
tending to equalize at higher energy transfer.

In Fig. \ref{fign3}, we compare DSF of superfluid gas with
corresponding DSF for normal fluid. The two curves do not show any
discernible shift of their peak values apparently due to
exponentially small gap. However, as $\omega$ decreases below the
value at which the maxima occurs, DSF in superfluid case exhibits
reducing gradient in contrast to that of normal case of almost
steady gradient.This feature may constitute an indication of  the
occurrence of BCS-type superfluidity in trapped Fermi gas. This
feature can be explained on the basis of inhomogeneous density
distribution of trapped gas. For a uniform
 Fermi superfluid,  in the single-particle excitation regime,
DSF remains zero until energy transfer exceeds $2\Delta$ at
 which it rises sharply
with the increasing energy transfer. For a superfluid trapped
Fermi gas, owing to the spatial distribution of the gap, DSF has a
structure below  $2\Delta_0$, where $\Delta_0$ represents the gap
at the trap center. As $\omega$ goes to zero, the gradient of
$S(\delta,{\mathbf q})$ vanishes. In the low energy regime
($\omega <2\Delta_0$), $S(\omega,{\mathbf q})$ varies with
$\omega$ nonlinearly. When $\omega$  approaches $2\Delta_0$, the
gradient changes abruptly implying a discontinuity (which may be
indiscernible experimentally on practical grounds). This behavior
can also be explained by considering the boundary condition
$2\Delta({\mathbf x}) < \omega$. This spatially dependent lower
bound on $\omega$ implies that, when $\omega$ is less than
$2\Delta_0$, the atoms at the central region of the trap can not
respond to the light fields via single-particle excitations, only
those atoms in the peripheral region can do so.

Fig. \ref{fign4} displays DSF for both the Case-I and Case-II for
larger scattering length $a_s = 3.21k_F^{-1}$ for which the gap is
$\Delta_0 = 0.15 \epsilon_F$. For both the Figs. \ref{fign3} and
\ref{fign4}, $q$ is fixed at $0.4k_F$. In comparison to the Fig.
\ref{fign3}, we notice that the peak of DSF in Fig. \ref{fign4}
exhibits a shift apparently due to the occurrence of relatively
larger gap. We further notice that the width has been broader with
peak value reduced by roughly one order of magnitude. This may be
attributed to the relatively larger interaction and hence larger
vertex correction.

\section{conclusion}

In conclusion we have presented a detailed theoretical analysis of
the response of Cooper-paired Fermi atoms due to off-resonant
light scattering at zero temperature. We have studied vertex
correction which is quite significant at low momentum. By making
use of the Zeeman shifts between two ground hyperfine spin states
and also between the excited state hyperfine spin manifolds, we
have shown that it is possible to transfer different amount of
momentum and energy to the two partner atoms of a Cooper pair.
Light polarization plays an important role in selective
single-particle excitations in superfluid Fermi atoms. Using
circularly polarized light in the presence of a magnetic field,
quasi-particle excitation can be obtained in one spin component
only.  We have analyzed the dynamic structure factor (DSF) due to
single-particle excitations under different physical conditions.
DSF shows a shift for large gap. In contrast to trapped normal
Fermi atoms, the gap inhomogeneity of trapped Cooper-paired Fermi
atoms  leads to relatively reduced gradient of DSF below
$2\Delta_0$, where $\Delta_0$ is the gap at the trap center. This
reducing gradient may constitute a signature of superfluid state.

Although the focal theme of this paper has been the theoretical
analysis of stimulated scattering of polarized light by superfluid
Fermi atoms under different physical conditions, there is  some
relevance of it in experiments with Fermi atoms. The question
arises how to detect experimentally spectrum of density
fluctuation (or DSF) of superfluid two-component Fermi atoms such
as $^6$Li. Towards this end, we wish to present some speculative
and suggestive discussions. We recall that the DSF of BEC has been
experimentally detected \cite{bragg,phonon} using stimulated light
scattering (or Bragg spectroscopy) and the well-established method
of time-of-flight measurements. One of the major difficulties in
evaluating DSF of two-component Fermi atoms from time-of-flight
measurements might stem from the fact that the initial information
about the momentum distribution of the atoms may be washed away
during expansion due to relatively large s-wave collisions of the
two hyperfine spin components. This difficulty can be circumvented
by the method of rapidly reducing the magnetic field (that induces
Feshbach resonance) just before switching off the trap as done in
numerous recent experiments \cite{solomon,ketterle,expt}. In light
scattering experiment, polarization-selective light scattering may
be useful in suppressing the collisions among the scattered atoms
during their expansion on being released from the trap. Since the
scattered atoms will be in a single spin state, there will be
diminished probability of collision among those atoms (the leading
order p-wave collision at low temperature is vanishingly small).
This may lead to better precision in time-of-flight spin-selective
measurements \cite{molecules,colorado} of scattered atoms. Order
of magnitude analysis of Ref. \cite{deb1} suggests that, with
large momentum transfer, it may be possible to distinguish the
scattered atoms in time of flight images. To reveal the
information about the momentum and density distribution of
scattered atoms, the time-of-flight images with and without Bragg
pulses should be compared. Furthermore, Bragg spectroscopy allows
one to choose different directions for scattered atoms, since the
scattering is of predominantly stimulated type. It may be possible
to scatter atoms in two opposite directions by using three or four
beam stimulated light scattering configuration as discussed in
\cite{deb1}. One can then explore the possibility of measuring the
correlation of two scattered atoms with opposite momentum by
similar technique as applied in recent theoretical \cite{theoshot}
and experimental \cite{jinshot} studies. Finally,
polarization-selective light scattering may be useful in exciting
BA mode the detection of which poses a challenging experimental
problem.

\appendix
\section*{Appendix}
\setcounter{section}{1}
 We here present an alternative derivation
of DSF  without vertex correction  using BCS-Bogoliubov mean-field
treatment. This DSF coincides with that obtained using leading
approximation as described in the text. It can  be defined by $
S({\mathbf q},\omega) = \sum_f | \langle f \mid
\sum_{\sigma=\uparrow,\downarrow}
\gamma_{\sigma\sigma}\rho_{\sigma}^{\dagger}({\mathbf q}) \mid 0
\rangle|^2 \delta(\omega- \epsilon_f + \epsilon_0) $ where
$|0\rangle$ represents the many-body ground state with energy
$\epsilon_0$ and the sum runs over all the final states
$|f\rangle$ which can be coupled to the ground state by the
density operator $\rho_{\sigma}({\mathbf q}) = \sum_{k}
a_{\sigma,\mathbf{k}+\mathbf{q}}^{\dagger}a_{\sigma, \mathbf{k}}$.
The DSF for the Case-I, that is, for the condition
$\gamma_{\uparrow\uparrow} = \gamma_{\downarrow\downarrow}$ has
been explicitly calculated in Ref. \cite{klein}. We here assume
$\gamma_{\uparrow\uparrow} \ne \gamma_{\downarrow\downarrow}$, and
as an extreme case we consider the case-II:
$\gamma_{\uparrow\uparrow} \simeq 0$ and
$\gamma_{\downarrow\downarrow}\rightarrow 1$. Setting
$\epsilon_0=0$, we can write
   \bearr S({\mathbf q},\omega) =
 \frac{V}{(2\pi)^3}\int d^3{\mathrm k}n_k(1-n_{k'}) \delta(\omega - E_{k'} - E_{k})
 \label{dsfu} \eearr
where $k' = |{\mathbf k} + {\mathbf q}|$ is the wave number of a
scattered atom, $V$ is the volume of the system and   $n_k = v_k^2
=  (1-\xi_k/E_k)/2$ is the momentum distribution function. Here
$E_k = \sqrt{\xi_k^2 + \Delta_k^2}$ is the energy of an elementary
excitation (Bogoliubov's quasiparticle),  $\Delta_k$ represents
the pairing gap and $\xi_k = \hbar^2k^2/(2m) - \mu$. Note that the
usual BCS coherence factor \cite{klein,schrieffer} $m({\mathbf k},
{\mathbf k}')=u_{k'}v_{k} + u_{k}v_{k'}$ where $u_k^2 = 1- v_k^2$,
has changed. This is due to the fact that the momentum and energy
transfer occurs in either partner of a Cooper pair because of
polarization-selective light scattering.

We here give an outline of the method of calculation of the
integral in Eq. (\ref{dsfu}). For notational simplicity, we denote
$\xi = \xi_k$ and $\xi'= \xi_{k'} \simeq \xi +
\mathbf{v}_k.\mathbf{p}_q$; and similarly we replace $E_{k}$ and
$E_{k'}$ by $E$ and $E'$, respectively. The integration  may be
restricted near $\xi = \epsilon_k -\mu \simeq 0$, since the
dominant contribution to the integration comes from $k$-values
near $  \epsilon_k \simeq \mu$. For convenience, we change the
variable of integration into $E$ by using the relation $d\xi =
EdE/(E^2 - \Delta^2)^{1/2}$. Using the identity \bearr
\delta(\omega - E-E') = \frac{\delta(E-E_0)}{\mid (1+dE'/dE)
\mid_{E=E_0}} \eearr where $E_0$ is the solution of the equation
$E + E' = \omega$, we have \bearr S({\mathbf q},\omega) =
\frac{V}{(2\pi)^3}\int d^3{\mathbf k} \delta(\xi_k) \left
[\frac{(E-\xi)(E' + \xi')}{4\mid 1+dE'/dE \mid \xi E'}\right
]_{E=E_0}. \eearr After a lengthy algebra as in Ref. \cite{klein},
we then obtain \bearr S({\mathbf q},\omega) = \frac{V}{(2\pi)^3}
\int d^3{\mathbf k} \delta(\xi_k) \frac{(\omega + {\mathbf
p}_q.{\mathbf v}_k)^2}{16\Delta^2\beta^3\sqrt{\beta^2-1}}, \eearr
where $\beta$ is defined in Eq. (\ref{beta}).  Let $z= \cos
\theta$, where $\theta$ is the angle between ${\mathbf k}$ and
${\mathbf q}$. Changing the variable of angular integration into
$x= v_kp_qz/\sqrt{\omega^2 - 4 \Delta^2}$, one obtains \bearr
S({\mathbf q},\delta) = \frac{N(0)\Delta^2}{2p_qv_F\omega}
\int_0^{x_0} dx \frac{(1 +
 jx^2)}{(1-jx^2)^{3/2}\sqrt{1-x^2}},   \label{sint}
\eearr where $j = 1- 4\Delta^2/\omega^2$ and $ x_0 = {\rm
Min}\left [1, \frac{p_q v_F}{(\omega^2 - 4\Delta^2)^{1/2}}\right
]$. In writing the above equation, we have considered the
weak-coupling case and hence replaced the chemical potential $\mu$
by the Fermi energy  $\epsilon_F$. If $2\Delta < \omega <
(p_qv_F)^2 + 4\Delta^2)^{1/2}$, then $x_0 = 1$ and the result is
\bearr S({\mathbf q},\delta) = \frac{N(0)\omega}{8p_qv_F} \left
[E(j) + M\hspace{0.05cm}_2F_1(3/2,3/2;2,j) \right ] \label{z11}
\eearr where $M = \pi j(1-j)/4$. Here $E(j)$ represents the
complete elliptic integral and $_2F_1(a,b;c,d)$ is the
hypergeometric function.  In the limit $\Delta \rightarrow 0$,
$E(j)\rightarrow 1$ and $M \hspace{0.05cm}_2F_1 \rightarrow 1$
leading to the result $S(\delta, {\mathbf q}) =
\nu_F\delta/(4p_qv_F)$ which is  half the dynamic structure factor
of normal fluid \cite{nozpines} consisting of noninteracting
quasi-particles within the specified energy range. The factor half
arises due to our initial assumption $\gamma_{\uparrow\uparrow}
\simeq 0$ and $\gamma_{\downarrow\downarrow} \rightarrow 1$ which
implies that only spin $\downarrow$ are scattered and since the
number of atoms in the two spin components  are assumed to be
equal, only half of the total number of atoms contribute to the
scattered flux of atoms.


\section*{References}
\begin{thebibliography}{10}

\bibitem{cooling} Cohen-Tannoudji C.  1998 {\it Rev. Mod. Phys.} {\bf 70}
 707; Chu S. 1998 {\it ibid} {\bf 70} 685 ; Philips W. D. 1998
{\it ibid} {\bf 70} 721

\bibitem{bec}  Anderson M.,  Ensher J. R.,  Matthews M. R.,
 Wieman C. E., and Cornell  E. A. 1995 {\it Science} {\bf 269} 198;
  Bradley C.
C.,  Sackett C. A., Tollett J. J. , and  Hulet R. G. 1995 {\it
Phys. Rev. Lett.} {\bf 75}  1687 ;  Davis K. B. ,  Mewes M. O.,
Andrews M. R., van Druten N. J.,  Durfee D. S., Kurn D. M. , and
Ketterle W. 1995 {\it Phys. Rev. Lett.} {\bf 75}  3969

\bibitem{einstein} Einstein A.  1924  {\it Sitzber. Kgl. Preuss. Akad.
Wiss.}
 261 1925 {\it ibid}  3

\bibitem{bose} Bose S. N. 1924 {\it Z. Phys.} {\bf 26} 178.

\bibitem{pred1} Stoof H. T. C., Houbiers M., Sackett C. A., and
Hulet R. G.  1996 {\it Phys. Rev. Lett.} {\bf 76} 10; M. Houbiers
{\it et al.} 1997 {\it Phys. Rev. A} {\bf 56} 4864 ; DeMarco B.
and Jin D. S. 1998  {\it Phys. Rev. A} {\bf 58} 4267

\bibitem{pred2} Heiselberg H. 2001 {\it Phys. Rev. A} {\bf 63}, 043606 ;
 H. Heiselberg, C. J. Pethik, H. Smith, and
Viverit L.  2000 {\it Phys. Rev. Lett.} {\bf 85} 2418; Bruun G. M.
and Heiselberg H. 2003 {\it Phys. Rev A} {\bf 65} 053407

\bibitem{jin} B. DeMacro and D. S. Jin 1999 {\it Science} {\bf 285} 1703

\bibitem{hulet} A. G. Truscott, K. E. Strecker, W. I. McAlexander, G.B.
Patridge, and R. G. Hulet 2001 {\it Science} {\bf 291} 2570

\bibitem{solomon} F. Schreck {\it et al.}  2001 {\it Phys. Rev. Lett.} {\bf 87}
080403;T. Bourdel {\it et al.} 2003 {\it ibid.} {\bf 91} 020402

\bibitem{thomas} S. R. Granade, M. E. Gehm, K. M. O'Hara, and J. E.
Thomas  2002 Phys.Rev.Lett. {\bf 88}, 120405;  O'Hara  {\it et
al.}  2002  {\it Science} {\bf 298}  2179

\bibitem{mit} Z. Hadzibabic {\it et al.},
 {\it Phys. Rev. Lett.} {\bf 88}, 160401 (2002).

\bibitem{italy} G. Roati, F. Riboli, G. Modungo, and M. Inguscio,
{\it Phys. Rev. Lett.} {\bf 89}, 150403 (2002).

\bibitem{grimm} Jochim S. {\it et al.} 2003  {\it Science} {\bf 302}
 2101

\bibitem{ketterle} Zwierlein M. W., Abo-Shaeer J. R., Schirotzek A.,
Schunck C. H., Ketterle W. 2005 {\it Nature} {\bf 435}  1047

\bibitem{gap1} Cin C. {\it et al.} 2004 {\it Science} {\bf 305}
1128

\bibitem{gap2} Greiner M., Regal C. A., and Jin
D. S. 2005 {\it Phys. Rev. Lett.} {\bf 94} 070403

\bibitem{duke} Kinast, J. {\it et al.} 2004  {\it Phys. Rev. Lett.} {\bf 92} 150402

\bibitem{innsbruck} Bartenstein M. {\it et al.} 2004 {\it Phys. Rev. Lett.} {\bf 92}
203201

\bibitem{stringari} Stringari S. 2004 {\it Europhys. Lett.} {\bf 65}
749

\bibitem{nozrink}  Nozi$\acute{e}$res  P. and  Schmitt-Rink S. 1985 {\it J.
Low. Temp. Phys.} {\bf 59}  195

\bibitem{randeria}  Sa de Melo C.A.R.,
Randeria M. and  Engelbrecht J.R. 1993 {\it Phys. Rev. Lett.} {\bf
71}  3202; Engelbrecht J.R., Randeria M. and Sa de Melo C.A.R.
1997 {\it Phys. Rev. B} {\bf 55} 15153

\bibitem{crossover}  Holland M.,  Kokkelmans S. J. J. M. F.,
Chiofalo M. L. and  Wasler R. 2001 {\it Phys. Rev. Lett.} {\bf 87}
 120406;  Timmermans E. {\it et al.} 2001 {\it Phys. Lett A} {\bf 285}
  228;  Ohashi Y. and  Griffin  A. 2002 {\it Phys. Rev. Lett.} {\bf 89}
  130402;  Hofstetter W. {\it et al. } 2002 {\it Phys. Rev. Lett.} {\bf
89} 220407

\bibitem{molecules}  Greiner M.,  Regal C. A. and  Jin D. S. 2003 {\it Nature} {\bf 426}
 537; Jochim S. {\it et al.}  2003 {\it Science} {\bf 302}  2101;
Zwierlein M. W. {\it et al.} 2003 {\it Phys. Rev. Lett.} {\bf 91}
 250401

\bibitem{expt}  Modugno G. {\it et al.} 2002 {\it Science} {\bf 297}
  2240;  Strecker K. E. {\it et al.} 2003 {\it Phys. Rev. Lett.}
{\bf 91} 080406;  Cubizolles J. {\it et al.} 2003 {\it Phys. Rev.
Lett.} {\bf 91}  240401

\bibitem{theory}  Falco G. M. and  Stoof H. T. C. 2004 {\it Phys. Rev. Lett.} {\bf 92}   130401;
 Carr L. D.,  Shlyapnikov G. V. and  Castin Y. 2004 {\it Phys. Rev.
 Lett.}
{\bf 92} 150404;  Heiselberg H. 2003 {\it Phys. Rev. A} {\bf 68}
 053616,  Perali A.,  Pieri P. and  Strinati G. C. 2003 {\it
Phys. Rev. A}, {\bf 68}  031601;  Perali A.,  Pieri P., Pisani L.
and  Strinati G. C. {\it lanl e-print cond-mat/0311309}.

\bibitem{zoller} T\"{o}rm\"{a} P. and  Zoller P. 2000 {\it Phys. Rev. Lett.} {\bf 85}  487;
 Bruun G. M. {\it et al.} 2001 {\it Phys. Rev. A} {\bf 64}
 033609;  Kinnunen J., Rodriguez M., and T\"{o}rm\"{a} P 2004 Phys. Rev. Lett. {\bf
 92} 230403; Bruun G. M. and Baym G. 2004 {\it Phys. Rev. Lett.} {\bf 93}
 150403; B$\ddot{u}$chler H. P.,  Zoller P., Zwerger W. 2004 {\it Phys. Rev. Lett.}, {\bf
 93}  080401

 \bibitem{theogap1} Kinnunen J., Rodriguez M., and T\"{o}rm\"{a} P
 2004
{\it Science} {\bf 305} 1131

\bibitem{huletp}  Zhang W.,  Sackett C. A. and  Hulet R. G. 1999 {\it Phys. Rev. A} {\bf 60} 504;
 Ruostekoski J. 1999 {\it Phys. Rev. A}  {\bf 60}  1775;
 Rodriguez M.
and T\"{o}rm\"{a} P. 2002 {\it Phys. Rev. A} {\bf 66} 033601.

\bibitem{mottelson} Bruun G. M. and Mottelson B. R. 2001 {\it Phys. Rev.
Lett.} {\bf 87}  270403

\bibitem{griffin} Ohashi Y. and Griffin A. 2003 {\it Phys. Rev. A} {\bf
67} 063612; Ohashi Y. and Griffin A.  {\it lanl archive
cond-mat/0503641}

\bibitem{minguzzi}  Minguzzi A.,  Ferrari G. and  Castin Y. 2001 {\it Eur.
Phys. J. D.} {\bf 17} 49

\bibitem{bamode} Bogoliubov N. N. 1958 {\it Nuovo Cimento} {\bf 7} 6;
Bogoliubov N. N., Tolmachev V. V., and Shirkov D. V. 1959 {\it A
New Method in the Theory of Superconductivity} (Consultants
Bureau, NY).

\bibitem{anderson} Anderson P. W. 1958 {\it Phys. Rev.}  {\bf 112} 1900

\bibitem{martin} Martin P. C., in 1969 {\it Superconductivity}, Vol.1, edited
by Parks R. D. (Dekker, NY).

\bibitem{bragg}  Stenger  J. {\it et al.} 1999 {\it Phys. Rev. Lett.}  {\bf 82}
 4569;  Ketterle W. and  Inouye S., {\it lanl e-print
cond-mat/0101424}.

\bibitem{abrikosov}  Abrikosov  A. A. and
Fal'kovski$\breve{i}$ L. A. 1961 {\it Soviet Physics JETP} {\bf
13}  179

\bibitem{klein}  Klein M. V. and  Dierker S. B. 1984 {\it Phys. Rev. B} {\bf 29}
 4976

\bibitem{nambu} Nambu Y. 1960 {\it Phys. Rev.} {\bf 117}  648

\bibitem{gorkov} Abrikosov A. A., Gorkov L. P., and Dzyaloshinski
1963 {\it Methods of Quantum Field Theory in Statistical Physics}
(Dover, NY).

\bibitem{sakurai}  Sakurai J. J. 1967 {\it Advanced Quantum Mechanics}
(Pearson Education, Inc.)

\bibitem{wilczek}  W. V. Liu and F. Wilczek 2003
{\it Phys.Rev.Lett.} {\bf 90} 047002

\bibitem{deb1}  Deb B.,  Mishra A.,  Mishra H. and  Panigrahi P.
K. 2004 {\it Phys. Rev. A}  {\bf 70}  011604

\bibitem{zwierlein}  Gupta  S. {\it et al.} 2003 {\it Science} {\bf 300}
 1723

\bibitem{schrieffer}  Schrieffer J. R. 1964 {\it Theory of
Superconductivity} ( W. A. Benjamin)

\bibitem{feynman} Feynman R. P. 1951  {\it Phys. Rev.} {\bf 84}  108

\bibitem{vaks} Vaks V. G., Galitskii, and Larkin A. I. 1962 {\it Soviet
Physics JETP} {\bf 14}  1177

\bibitem{analyt}  Marini M.,  Pistolesi F. and  Strinati G. C. 1998  {\it Eur.
Phys. J. B } {\bf 1}   151

\bibitem{ward} Ward J. C., 1950 {\it Phys. Rev.} {\bf 78}  182

\bibitem{nozpines}  Pines D. and  Nozi$\grave{e}$res P. 1989 {\it The
Theory of Quantum Fluids},  Vol.{\bf I} (Addison-Wesley)

\bibitem{deb2} Deb B. {\it LANL e-print archive Cond-mat/0505692}

\bibitem{mancini} Angelis G. F. D. and Mancini F. 1974 {\it Nuovo
Cimento} {\bf 10}  654

\bibitem{ambe} Amvbegaokar V. and Kadanoff L. P. 1961 Nuovo Cimento
{\bf 22} 914

\bibitem{houbiers}   Houbiers M. {\it et al.} 1997 {\it Phys. Rev. A} {\bf
56}  4864;  Vichi L. and  Stringari S. 1999 {\it Phys. Rev. A}
{\bf 60}   4734

\bibitem{strinati} Perali A. Pieri P. and Strinati G. C. 2003 {\it
 Phys. Rev. A} {\bf 68}  031601

 \bibitem{pitaevskii}  Viverti L.,  Giorgini S.,  Pitaevskii L. P. and
 Stringari S. 2004 {\it Phys. Rev. A} {\bf 69}   013607

\bibitem{perali} Perali A. Pieri P., Pisani L.  and Strinati G. C. 2004 {\it
 Phys. Rev. Lett} {\bf 92} 110401

 \bibitem{colorado} Regal C. A.,  Greiner M. and  Jin D. S. 2004 {\it Phys. Rev.
Lett.} {\bf 92}   040403;  Zwierlein M. W.,  Stan C. A., Schunck
C. H.,  Raupack S. M. F.,  Kerman A. J. and  Ketterle W. 2004 {\it
Phys. Rev. Lett.} {\bf 92} 120403

\bibitem{phonon}  Stamper-Kurn D. M. {\it et al.} 1999 {\it Phys. Rev.
Lett.} {\bf 83} 2876; Vogels J. M. {\it et al.} 2002 {\it Phys.
Rev. Lett.} {\bf 88} 060402

\bibitem{theoshot} Altman E. Demler E., Lukin M. D. 2004 {\it Phys. Rev. A}
{\bf 70}  (2004) 013603

\bibitem{jinshot} Greiner M., Regal C. A., Stewart J. T. and Jin D. S. 2005 {\it
Phys. Rev. Lett.} {\bf 94} 110401

\end{thebibliography}
\end{document}